\documentclass[usenatbib]{mnras}

\usepackage{newtxtext,newtxmath}

\usepackage[T1]{fontenc}
\usepackage{ae,aecompl}

\usepackage{amssymb,multirow,tabularx}
\usepackage{amsmath}
\usepackage{graphicx}
\usepackage{BibDef}
\usepackage{xcolor}
\usepackage[normalem]{ulem}
\usepackage{hyperref}
\hypersetup{
	colorlinks=true,        
	linkcolor=blue,         
	citecolor=blue,         
}

\newcommand{\msun}{~\rm M_{\large \odot}}

\title[NSB orbits and r-process enrichment]{Neutron star binary orbits in their host potential: effect on early r-process enrichment}

\author[M. Bonetti et al.]{
Matteo Bonetti,$^{1,2,3}$\thanks{E-mail: matteo.bonetti@unimib.it}
Albino Perego,$^{3,4}$\thanks{E-mail: albino.perego@unitn.it}
Massimo Dotti$^{2}$
and Gabriele Cescutti$^{5}$
\\
$^1$DiSAT, Universit\`a degli Studi dell'Insubria, Via Valleggio 11, 22100 Como, Italy\\
$^2$Dipartimento di Fisica ``G. Occhialini'', Universit\`a degli Studi di Milano-Bicocca, Piazza della Scienza 3, 20126 Milano, Italy\\
$^3$INFN, Sezione di Milano-Bicocca, Piazza della Scienza 3, 20126 Milano, Italy\\
$^4$Dipartimento di Fisica, Universit\`a degli Studi di Trento, via Sommarive 14, 38123 Trento, Italy\\
$^5$INAF, Osservatorio Astronomico di Trieste, Via G.B. Tiepolo 11, I-34143 Trieste, Italy\\
}

\date{Accepted 2019 September 9. Received 2019 September 9; in original form 2019 May 28}

\pubyear{2019}

\begin{document}
\label{firstpage}
\pagerange{\pageref{firstpage}--\pageref{lastpage}}
\maketitle

\begin{abstract}
Coalescing neutron star binary (NSB) systems are primary candidates for $r$-process enrichment of galaxies. The recent detection of $r$-process elements in ultra-faint dwarf (UFD) galaxies and the abundances measured in classical dwarfs challenges the NSB merger scenario both in terms of coalescence time scales and merger locations.
In this paper, we focus on the dynamics of NSBs in the gravitational potentials of different types of host galaxies and on its impact on the subsequent galactic enrichment. We find that, for a $\sim t^{-1}$ delay time distribution, even when receiving a low kick ($\sim 10~{\rm km~s^{-1}}$) from the second supernova explosion, in shallow dwarf galaxy potentials NSBs tend to merge with a large off-set from the host galaxy. This results in a significant geometrical dilution of the amount of produced $r-$process elements that fall back and pollute the host galaxy gas reservoir. The combination of dilution and small number statistics produces a large scatter in the expected $r$-process enrichment within a single UFD or classical dwarf galaxy.
Comparison between our results and observed europium abundances reveals a tension that even a systematic choice of optimistic parameters in our models cannot release. Such a discrepancy could point to the need of additional $r$-process production sites that suffer less severe dilution or to a population of extremely fast merging binaries.
\end{abstract}




\section{Introduction}

The merger of compact binaries comprising at least one neutron star (NS) has 
long been thought to be the site for the production of a significant fraction
of the heavy elements above the iron group via the so called $r$-process nucleosynthesis \citep[e.g.][]{Lattimer1974,Eichler1989,Freiburghaus1999a}. 
The recent detection of a kilonova transient (AT2017gfo)
associated with the gravitational wave (GW) signal produced by two NSs in the late phases of their inspiral \citep[GW170817, e.g.][]{Abbott2017,Abbott2017f,Drout2017,Tanaka2017,Pian2017,Tanvir2017,Kasen2017,Nicholl2017,Chornock2017} has finally provided strong observational support to these ideas and confirmed that NSB mergers are one of the major (if not the main) production site for $r$-process nucleosynthesis elements \citep[see e.g.][]{Thielemann.etal:2017,Rosswog2017a,Hotokezaka2018,Cote2018a}.

The unprecedented quality of the kilonova detection has provided a glimpse
of the potential variety associated with this new class of transients. 
The  optical  and  infrared  electromagnetic data are 
well explained by the radioactive decay of $\sim 0.05 \msun$ of material 
\citep[e.g.][]{Cowperthwaite2017,Rosswog2017a,Tanaka2017,Tanvir2017,Smartt2017,Kasen2017,Perego2017a}.
The presence of different peaks in the light curves and the spectral evolution 
can be modelled by different components in the outflows
with different compositions and, possibly, different physical origins.
Modelling of the matter outflow and of its properties is presently accomplished
by numerical simulations of the merger and of its aftermath.
Matter expelled within the first milliseconds after the NS collision is dubbed dynamical ejecta
\citep[e.g.][]{Korobkin2012,Bauswein2013a,Hotokezaka2013a,Wanajo2014,Sekiguchi2015,Radice2016,Bovard2017,Radice2018,Radice2018b}.
On longer time scales (a few hundreds milliseconds) a significant fraction of the ejecta can be expelled as baryonic winds that originate from the merger remnant
\citep[e.g.][]{Fernandez2013,Metzger2014,Perego2014b,Siegel2014,Just2015,Martin2015,Siegel2018,Radice2018c}. 
This ejecta can come both from the disc that forms around the central remnant 
(either a massive neutron star or a black hole) and from the central 
remnant itself, as long as it does not collapse to a black hole.

Observations of $r$-process elements in the atmosphere of metal-poor stars
in our galaxy and in nearby classical dwarf galaxies hint at the occurrence of $r$-process nucleosynthesis also in metal-poor environment, corresponding to the very early 
stages of the galaxy evolution \citep{Mcwilliam1998,Sneden2003,Shetrone2003,Honda2006,Francois2007,Sneden2008,Roeder.etal:2014,Ural2015,Jablonka2015,Hill2018}.
Moreover, the abundance of europium (an element synthesised mainly by $r$-process
nucleosynthesis) presents a large scatter 
at very low metallicities (${\rm [Fe/H]} < -3$), suggesting that 
$r$-process elements must be synthesized in rare and isolated events that inject 
a significant amount of heavy elements into a relatively small amount of gas \citep[e.g.][and references therein]{Sneden2008}. 
This rare-event/high-yield scenario is also corroborated by the comparison of iron and plutonium abundances in deep-sea sediments \citep{Hotokezaka2015}. This hypothesis was also tested in stochastic and inhomogeneous chemical evolution models \citep{Cescutti15,Wehmeyer2015}, which succeed to explain the chemical spread of r-process abundances in halo stars, but only assuming very short delay for the NSB merger.

Among the smallest dwarfs, called ultra-faint dwarf galaxies (UFD), Reticulum II \citep{Ji2016,Roederer2016,Ji2016a}
and very likely also Tucana III \citep{Drlica2015,Simon2017,Hansen2017,Marshall2018} show
a large excess of $r$-process elements, while for all the other UFDs robust upper limits on their $r$-process element abundances have been set \citep{Frebel2010A,Simon2010,Koch2013,Francois2016}. The presence of $r$-process elements in UFDs poses 
serious challenges to NSB mergers as origin of the $r$-process nucleosynthesis
elements. First, these galaxies have low escape velocities 
\citep{Walker2015}. The kick imparted to any newly born NS by its 
CCSN explosion could potentially eject the NS from the galaxy, even 
when the stellar binary system survives the second supernova
explosion. Second, their old stellar
population is thought to be the results of a fast star formation episode, that ends within the first Gyr of the galaxy evolution after the first CCSN explosions expel a significant fraction of baryons \citep{Brown2014,Weisz2015}.

A firm understanding of the $r$-process enrichment in dwarf galaxies is also essential to understand metal poor stars in the Milky Way halo. Indeed both observations \citep{Frebel2010,Ivezic2012} and theoretical models \citep{Helmi2008,Griffen2016} point to the fact that the dwarf satellite galaxies that we observe nowadays around our Galaxy are the remnants of a large population of dwarfs that long ago merged with it to form the galactic halo stellar population. Moreover, recent results coming from GAIA \citep{GAIA} point to the fact that a large fraction of the Milky Way halo was actually formed through a merger with a single and relative massive satellite \citep{Haywood2018,Helmi2018}.

Assuming that fast mergers of compact binary systems require a high natal kick, \citet{Bramante2016} disfavoured NSB mergers as the source of the $r$-process material observed in Reticulum II. In contrast, further studies concerning the observed distributions of the orbital parameters of double NS systems \citep{Beniamini2016a} and the formation channels of such systems \citep{Tauris_et_al_2017} suggest that a large fraction of NSB systems could have received a rather small kick and have ejected a small amount of mass as a consequence of the second CCSN explosion.
These conclusions imply that a large fraction (up to 60\%) of double NS systems could be retained even by UFD galaxies and to merge within the first Gyr of galaxy evolution \citep{Beniamini.etal:2016}.
Also \citet{Safarzadeh2019a} reached opposite conclusions with respect to \citet{Bramante2016} by considering the possibility that high-kick NSBs are either on highly eccentric orbits or form with very short separations due to an additional mass-transfer between the first-born neutron star and a naked helium star, progenitor of the second neutron star.

Due to the low stellar content of UFD galaxies and the subsequent small number of NSB systems expected in these galaxies, objects like Reticulum II and Tucana III should be a minority. Moreover, in this scenario these UDFs should have hosted a single NSB merger that was able to significantly enrich them in $r$-process material.
These first analyses was later refined by \citet{Beniamini2018}, who considered that some of the $r$-process material synthesised during the mergers could still escape from the galaxy, thanks to the large kinetic energy that characterizes NSB merger ejecta ($\sim 10^{50}$~erg). By performing a more detailed analysis of the enrichment in iron from CCSNe and in $r$-process elements from NSB mergers in UFD galaxies, they confirmed the compatibility between the abundances observed and an $r$-process enrichment due to the ejection of material from rare events taking place inside dwarf galaxies with low escape velocities.

In many of the above mentioned studies, it was assumed that a NSB bound to its host coalesces always well inside the galaxy. The potential relevance of the merger location, and thus of the imparted kick, in explaining the abundances observed in UFDs was first underlined by \citet{Safarzadeh2017}. In particular, if the merger time is not extremely short \citep{Bonetti2018,Safarzadeh2019a}, the binary will start orbiting the galaxy and there is a high chance that the merger will still happen far from the regions of the galaxy where the next generation of stars will form. In this work, we systematically explore the potential effect of the dilution of the ejecta on enrichment of the $r$-process material due to the location of NSB mergers relative to the host galaxy \citep[see also][for a similar analysis focused on the Milky-Way]{Safarzadeh2017b}. Under a common set of minimal assumptions about the properties of the host galaxy, the star formation rate inside it, the NSB birth and coalescence as well as the ejecta properties, we investigate a wide sample of galaxy masses and we compute the fraction of the ejecta material retained by the galaxy.

The paper is structured as follows: in sections~\ref{sec:method}-\ref{sec:retained ejecta} we present the model we have adopted for our calculations and we detail all its components. The results we have obtained are presented in section~\ref{sec:results}, while in section~\ref{sec:discussion} we discuss our results and compare them with observations. We finally conclude in section~\ref{sec:conclusions}. 

\section{Overview of the chemical enrichment model}
\label{sec:method}

Modelling the evolution of galaxies and of their chemical enrichment 
is an extremely complex task, since it requires to follow many different processes that span a huge range of scales (both in space and in time). In this section, we first present a summary of the (simplified) model adopted in this work to study the enrichment in $r$-process material due to NSB mergers. 

We consider disc galaxies with a different content of baryonic matter $M_{\rm b}$, ranging from $10^5 M_{\odot}$ up to $10^8 M_{\odot}$, as well as a model with $5 \times 10^{10} M_{\odot}$. The former interval is expected to correspond to the initial gas content of ultra-faint and classical dwarf galaxies while
the latter value is the one of a MW-like galaxy.
During the cosmic history, gas is converted into stars with a certain star
formation rate ($f_{\rm SFR}$) starting from the galaxy formation ($t = 0$) up to 
$t \approx T_{\rm SF}$.
For a MW-like galaxy, $T_{\rm SF} \sim 14~{\rm Gyr}$ and the final stellar content is thought to be comparable to the initial gas mass. For smaller galaxies SN feedback and/or environmental processes (e.g. tidal perturbations and ram pressure stripping) are expected to quench the conversion of gas into stars on shorter timescales and to remove a fraction of the gas from the galaxy \citep{Revaz2018}. In the case of UFD, we assume $T_{\rm SF} \lesssim 1~{\rm Gyr}$, while for classical dwarfs $T_{\rm SF} \sim 3-5~{\rm Gyr}$.
The SFR has a non-trivial dependence on the cosmic time and on the individual history of each galaxy. Since we are not interested in the detailed time evolution of the metal content nor in the reproduction of a specific galaxy model, we adopt an exponential dependence on time, possibly dependent on the initial gas mass. The stellar content at any time $t$, $M_{*}(t)$ is then computed as the integral of $f_{\rm SFR}$ over time.

For each galaxy, we generate a pool of $N$ initial conditions for NSBs forming from stellar binaries as a consequence of a double CCSN explosion.
For each NSB, the relevant initial conditions after the second SN are the initial position in the galaxy, the NS masses ($m_1,m_2$), the semi-major axis and eccentricity ($a,e$), and the center of mass (CoM) velocity of the NSB with respect to galaxy frame ($\mathbf{V}_{\rm CM}$). 
For all the dwarf galaxies in the our sample, the expected number of merger is
$\lesssim 10^4$, therefore we choose $N \gg N_{\rm merg}$ to properly span the parameter space. 
For the MW-like galaxy, the number of mergers could potentially exceed a few millions. We expect such a high number of configurations to be large enough to significantly cover the whole NSB parameter space. Thus we consider a pool of $N \sim 2 \times 10^6$ NSB configurations. We evolve each initial condition by integrating the trajectory of the CoM of the NSB for the GW driven coalescence time $T_{\rm gw}(a,e)$, unequivocally determined by the initial binary parameters. Thus, we can associate to each NSB in our pools the time at which the merger happens and the corresponding location.

To compute the amount of $r$-process material produced by NSB mergers that has enriched a specific galaxy at a time $t$ we proceed as follows:
\begin{itemize}
    \item starting from the galaxy stellar mass at $t$, $M_{*}(t)$, we estimate the amount of CCSN explosions, $N_{\rm CCSN}$, and from that the expected number of stellar binaries surviving a double CCSN explosion and producing a NSB system within $t$, $N_{\rm merg}$;
    \item we sample the actual number of NSB mergers $N_r$ from a Poisson distribution with average equal to $N_{\rm merg}$, and we randomly choose $N_r$ cases from the galaxy NSB pool. We stress here that a (possibly large) fraction of these binaries could merger on a time $T_{\rm gw} \lesssim t$. In the case of the MW-like galaxy, if 
    $N_r > N_{r,{\rm max}} = 10^5$ we select $N_{r,{\rm max}}$ NSB from the pool of 
    $N \sim 2 \times 10^6$ elements. In this case, at the end of the analysis, the true values of the total and retained masses are obtained by rescaling the computed quantities by a factor $N_{r}/N_{r,{\rm max}}$. This approach is motivated by the need of reducing the necessary computations and the amount of data. However, it is justified by the large number of expected mergers with respect to their intrinsic variability;
    \item for each sampled binary merging within $t$, we model the properties of the material ejected during the coalescence, and its expansion due to the interaction with the diffuse gaseous halo of the host in order to determine the fraction and composition of the ejecta that gets injected in the galaxy disc. We consider that the ejected mass can be in form of wind ejecta (isotropically distributed) as well as dynamical ejecta, characterized by both a polar and an equatorial component.
\end{itemize}

Finally, in order to estimate both the average and the scatter in the distribution of the total ejected and retained material, for each galaxy model we perform 200 different realizations of the expected NSB populations. For each realization we assume full mixing of the ejecta. Despite being a simplification, this hypothesis is not a limitation because we will compare with estimates of the total galactic abundances. More detailed studies assuming inhomogeneous mixing and stochastic stellar enrichment are planned for the future.

In the following we fully detail the procedure adopted in our model.
In section~\ref{sec:potential}, we start with the modeling of the host galaxy and with 
the determination of the number of stars, SNae and merging NSBs 
for each galaxy model. We also discuss the modelling of the host galaxy potential, which determines the velocity of the NSB progenitor binary system, its CoM orbital evolution as well as the escape velocity.
We then discuss the initialization of the intrinsic parameters of each NSB (section~\ref{sec:internal}): the masses of the two NSs ($m_1$ and $m_2$), the semi-major axis and eccentricity of the the binary ($a$ and $e$), and the kick that the NSB gets due to the two SNae explosions ($\mathbf{V}_{\rm CM}$). 
The initialization of the initial position and velocity of the NSB CoM is described in section~\ref{sec:external}.
The ejecta properties immediately after the NSB coalescence are described in section~\ref{sec:ejecta1}, while their evolution within the host halo is described in section~\ref{sec:ejecta2}.

\section{Galaxy models}
\label{sec:potential}

We consider five different disc galaxies spanning a wide range of possible masses, $M_{\rm b}$: four dwarf galaxy models with total baryonic mass of $10^5\msun$, $10^6\msun$, $10^7\msun$, and $10^8\msun$ respectively, and a MW-like host modelled with five dynamical baryonic components \citep[galactic bulge + thin and thick stellar disks + HI and H$_2$ disks, see next and][for full details]{Barros2016} surrounded by a dark halo. 

The stellar population of UFD galaxies seem to be dominated by very old ($\sim 12~{\rm Gyr}$) stars \citep[][and references therein]{Brown.etal:2014}, pointing to a rapid SF, $\lesssim 1~{\rm Gyr}$. Old stars ($\lesssim 10~{\rm Gyr}$) dominate also dwarf galaxies \citep{Grebel:1997}, indicating that the bulk of the SF in them happens within 2-3~Gyr.
To approximately catch the dependence of the duration of the star formation on the initial baryonic mass $M_{\rm b}$, we assume a power-law dependence normalized to the MW-like case:
\begin{equation}
    T_{\rm SF}(M_{\rm b}) = T_{\rm SF}(M_{\rm MW}) \left( \frac{M_{\rm b}}{M_{\rm MW}} \right)^{\alpha} 
\end{equation}
with $\alpha = 0.2$ and $T_{\rm SF}(M_{\rm MW})=14~{\rm Gyr}$. For the star formation rate, we adopt the following exponential dependence:
\begin{equation}
    f_{\rm SFR}(t,M_{\rm b}) =
    \begin{cases}
    A~M_{\rm b}~\exp{\left( -\frac{t}{\tau(M_{\rm b})} \right)} & \qquad {\rm if}~t < T_{\rm SF} \\
    0 & \qquad {\rm otherwise} \, .
    \end{cases}
\end{equation}
where $A$ is a constant fixed by the requirement that the MW-like model reproduces
the presently observed star formation rate in the MW, i.e.
$f_{\rm SFR}(10~{\rm Gyr},M_{\rm MW}) = 1.65~\msun~{\rm yr^{-1}}$ \citep{Liquia2015}.
For the timescale appearing inside the exponential factor, 
we choose $\tau = T_{\rm SF}/2$. This choice 
is broadly compatible with (simple) models of the MW \citep[e.g.][]{Snaith2014} 
and of classical dwarf galaxies \citep{North2012}.
Finally, the gas mass converted into stars as a function of time and of initial baryonic mass
can be easily computed as 
\begin{equation}
    M_{*}(t,M_{\rm b}) =
    \begin{cases}
    A~M_{\rm b}~T_{\rm SF} \left( 1-e^{-2t/T_{\rm SF}} \right)/2 & \qquad {\rm if}~t < T_{\rm SF} \, , \\
    A~M_{\rm b}~T_{\rm SF} \left(1-e^{-2}\right)/2 & \qquad {\rm otherwise} \, .
    \end{cases}
\end{equation}

\subsection{Dwarf galaxies}
\label{sec:dwarf}

We assume that the potential of the host is well described by only two components: a baryonic disc (where we assume that the stars and gas follow the same profile) and a dark matter spherical halo.

The baryonic density profile of the galaxy is modelled as an exponential disk:
\begin{equation}\label{eq:disc}
	\rho_d(R,z) = \dfrac{M_{\rm b}}{4 \pi R_d^2 z_d} \exp{\left(-\frac{R}{R_d}\right)}\; {\rm sech}^2\left(\frac{z}{z_d}\right)  
\end{equation}
where $R$ and $z$ are the cylindrical radial and vertical coordinates, while the length scales $R_d$ and $z_d$ are \citep{Mo1998}
\begin{align}
    R_d &= 0.7 \left( \dfrac{M_b}{10^8 M_{\odot}} \right)^{1/3} {\rm \ kpc},\\
    z_d &= 0.2 R_d.
\end{align}
An analytic form for the potential (and consequently for the acceleration) cannot be obtained, we therefore employ a numerical sampling of the disk density profile with $\approx 8\times 10^5$ tracers, and splitting every tracer in 8 sub-tracers by changing the sign of 1, 2 or all the 3 coordinates, in order to preserve the symmetry of the potential. 
The disc acceleration $\mathbf{a}_{\rm disc}(\mathbf{r})$ is then evaluated through direct summation over all the sampled particles, where we use a gravitational softening of $\epsilon_{\rm soft} = 0.01 \left(M_b/10^{8} \msun\right)^{1/3}$ kpc to avoid spurious strong scattering due to the finite number of tracers used \citep[see e.g.][]{Monaghan1985}.
Such a kind of exponential disk profiles\footnote{Or equivalently S\'ersic profiles with a very small S\'ersic index} fits well the stellar brightness profile of many observed dwarf galaxies, including dwarf ellipticals and dwarf spheroidals, \citep[see e.g.][]{Faber.Lin:1983,Binggeli.etal:1984,Kormendy:1985,Graham:2002,Graham.Guzman:2003}. Furthermore, it implies that the stellar dynamics is dominated by rotation. While this seems to be the case for the vast majority of dwarf galaxies \citep{Kerr.etal:1954,Kerr.deVau:1955,Swaters.etal:2009}, for dwarf spheroidal galaxies with less rotational support and embedded in virialized galaxy clusters it is reasonable to assume that they were more rotationally supported in the past (during their star formation epoch) and lost their coherence due to galaxy harassment \citep{Moore.etal:1996}.

The dark matter density is assumed to follow a NFW profile:
\begin{equation}
    \rho_{\rm DM}(r) = \dfrac{\rho_0}{\dfrac{r}{r_h} \left(1 + \dfrac{r}{r_h}\right)^2},
\end{equation}
where the normalization $\rho_0$ is given by
\begin{equation}
    \rho_0 = \dfrac{M_{DM}}{ 4\pi r_h^3 \left( \log(1+C) - \dfrac{C}{1+C} \right) },
\end{equation}
with $M_{\rm DM}=  100 M_b$,\footnote{Dwarf galaxies are among the structures with the smallest baryonic to dark matter ratios. Detailed studies \citep[e.g.][]{McGaugh.etal:2010,Chan:2019} suggest a weak dependence on the baryonic mass, $\sim M_{b}^{0.23}$. For simplicity in our study we assume one single value broadly compatible with the expected ratio in dwarf galaxies.} $C=9.4$ and the scale radius is given by:
\begin{equation}
    R_h = 32.9 \left(\dfrac{M_b}{10^8 M_{\odot}}\right)^{1/3} \ \rm kpc.
\end{equation}
The potential generated by such distribution is analytic
\begin{equation}
	\Phi_{\rm DM}(r) = -\dfrac{4\pi \rho_0 G r_h^3}{r} \ln\left(1+\dfrac{r}{r_h}\right),
\end{equation}
which allow us to compute the acceleration directly from 
\begin{equation}
    \mathbf{a}_{\rm DM}(\mathbf{r}) = -\nabla \Phi_{\rm DM}(r).
\end{equation}
The total acceleration is then computed as $\mathbf{a}=\mathbf{a}_{\rm DM}+\mathbf{a}_{\rm disc}$.

\subsection{MW-like galaxies}
Following \citet{Barros2016}, we model the MW by considering a dark halo, a galactic bulge plus two stellar disks (thin and thick) and two gaseous disks (HI and H$_2$). The dark halo is modelled as a logarithmic potential, while the bulge is assumed spherically symmetric and described by an Hernquist profile \citep{Hernquist1990}. The description of the baryonic disks requires instead additional modelling. In fact, despite from an observational point of view these disks can be accurately fitted assuming exponential disks, as pointed out in section~\ref{sec:dwarf}, they do not admit any analytical form for the gravitational potential nor for the acceleration. This does not represent a real issue for the dwarf galaxy case because the global number of NSB that we have to simulate is rather small. On the contrary, for a MW-like galaxy this number can exceed several millions, with a severe impact on the performance of our calculation. This motivated our choice to describe the baryonic disks with analytical potential-density pairs. In particular, in \citet{Barros2016}, each of the four disk is modelled with a superposition of three Miyamoto-Nagai (MN) disks \citep{Miyamoto1975}, where one of the three mass parameter is usually negative in order to mimic the sharp decrease that characterise the density profile of exponential disks. We address the interested reader to section 2 of \citet{Barros2016} for the detailed description of the model, and to table 3-4 in the same paper for the values of the parameter that best describe the MW and that we implemented in our model.

\subsection{Average number of NSB mergers for a given galaxy}
\label{sec:mean_mergers}

\begin{table*}
\begin{tabular}{ | c | c | c | c | c | c | c | c | c  c |  }
\hline
Model & $M_{\rm b}$ & $M_{\rm DM}$ & $T_{\rm SF}$ & $M_{*}(T_{\rm SF})$ & $R_{\rm d}$ & $z_{\rm d}$ & $N_{\rm CCSN}(T_{\rm SF})$ & \multicolumn{2}{c|}{$N_{\rm NSB}(T_{\rm SF})$} \\
{~} & $[\msun]$ & $[\msun]$ & [Gyr] & $[\msun]$ & $[{\rm kpc}]$ & $[{\rm kpc}]$ &  & min & max \\ \hline
Dwarf 1 & $10^5$ & $10^7$    & 1.0 & $6.0 \times 10^3$ & 0.07 & 0.014 & 56 & 0.04  & 1.12 \\
Dwarf 2 & $10^6$ & $10^8$    & 1.6 & $9.5 \times 10^4$ & 0.15 & 0.030 & 886 & 0.65 & 17.7\\
Dwarf 3 & $10^7$ & $10^9$    & 2.5 & $1.5 \times 10^6$ & 0.32 & 0.065 & $1.40\times 10^4$ & 10.4 & 281 \\
Dwarf 4 & $10^8$ & $10^{10}$ & 4.0 & $2.4 \times 10^7$ & 0.7  & 0.14  & $2.22\times 10^5$ & 164 & $4.44 \times 10^3$\\
\hline
\end{tabular}
\caption{Summary of the most relevant properties of the different dwarf galaxy models employed in our study (full details about MW-like parameters can be found in \citet{Barros2016}, tables 3-4). $M_{\rm b}$ and $M_{\rm DM}$ are the baryonic and dark matter masses, $T_{\rm SF}$ is the duration of the star formation, $M_*(T_{\rm SF})$ is the mass converted in star at $T_{\rm SF}$, $R_{\rm d}$ and $z_{\rm d}$ the scale radius and height of the disk, $N_{\rm CCSN}(T_{\rm SF})$ the average number of CCSN exploded by $T_{\rm SF})$, while the two values of $N_{\rm NSB}$ represent the minimum and the maximum average number of NSB formed.
The least (most) optimistic case is obtained considering one NSB every 50 (1350) CCSNe.}
\label{table: model summary}
\end{table*}

Given the large uncertainties that affect the determination of this value, we adopt a fairly simple but physically motivated approach. 
We assume a standard stellar initial mass function (IMF), identical for all models \citep{Kroupa2001}. For a given galaxy model and a given time, we compute the number of stars that have exploded as CCSNae ($N_{\rm CCSN}$)
as the number of stars with mass greater than $8\msun$. In doing that, we are
implicitly assuming that the evolution timescale of massive stars is much shorter than any time we are going to explore. We then assume that the number of NSB systems that form represents a fraction of the total number of SNe that have exploded. In particular, we parametrise $N_{\rm NSB}$ as
\begin{equation}
    N_{\rm NSB} = N_{\rm CCSN}/x
\end{equation}
where for $x$ we choose four values logarithmically distributed between an optimistic and a very pessimistic estimate, i.e. $x = [50,150,450,1350]$. 
This broad interval covers present uncertainties in the determination of the ratio between the number of exploding
CCSNe and the number of forming NSB systems, as obtained in detailed population synthesis models \citep{Giacobbo2018}.

In Table~\ref{table: model summary} we summarize the properties of the different dwarf galaxy models, alongside their names used in the following.


\section{Binary NS and their ejecta}

\subsection{NSB internal parameters and kick velocity}
\label{sec:internal}

Both the masses and the semi-major axes of each binary are sampled from observationally constrained distributions
\citep[e.g.][and references therein]{Tauris_et_al_2017}. The masses of the two NSs $m_1$ and $m_2$ are randomly sampled from normal distribution with $\mu_1 = 1.4 \msun$, $\mu_2 = 1.34 \msun$ and $\sigma = 0.14\msun$. The semi-major axis $a$ is evaluated from a log-uniform distribution where upper and lower limit are set selecting twice the maximum and half the minimum of ``observed'' NS binary orbital periods, i.e. $P \in [0.05,100]$ days.\footnote{Assuming a total mass of $\sim 2.8\msun$ these limits correspond to a minimum and maximum semi-major axis of about $\sim 0.8$ and $\sim 130$ solar radii.}

Tidal circularization acting after the first SN explosion does not allow to constrain the kick velocity experienced by the first-born NS. As a consequence, we sample the first kick experienced by the binary CoM ($\mathbf{V}_{\rm CM,1}$) assuming isotropy and a uniform magnitude distribution between 10 and 20 km/s, as constrained by the observed velocities of High-Mass X-Ray Binaries \citep[see e.g.][]{Coleiro2013}.\footnote{For this reason, we do not need to consider the effect of the mass lost during the first SN.}
	
The kick velocity associated to the second SN ($\mathbf{V}_{\rm kick,2}$) is isotropically generated in the $m_2$ rest frame. The kick magnitude ($V_{\rm kick,2}$) is distributed as follows: following \citet{Beniamini2016a}, we assume that the distribution is bimodal, with 60-70$\%$ of the cases following a log-normal "low-kick distribution" 
\begin{equation}
    p_{l}(V_{\rm kick,2}) = \dfrac{1}{\sqrt{2 \pi} \ V_{\rm kick,2} \ \sigma_{\ln{V},l}} \exp\left(\frac{\ln({V_{\rm kick,2}/\bar{V}_{l}})^2}{2\sigma_{\ln{V},l}^2}\right), 
\end{equation}
with $\bar{V}_{l}=9$ km/s and $\sigma_{\ln{V},l}=0.8$, and the remaining 40-30$\%$ of the cases are sampled from a second log-normal distribution $p_{h}(V_{\rm kick,2})$ with the corresponding parameters $\bar{V}_{h}=158$ km/s and $\sigma_{\ln{V},h}=0.5$.\footnote{The value of the chosen parameters are selected in order to broadly reproduce the merger time and the eccentricity distribution of observed NSB.}

The mass loss associated to the second SN is also sampled from a log-normal distribution:
\begin{equation}
    p_{l,h}(\Delta M) \dfrac{1}{\sqrt{2 \pi} \ \Delta M \ \sigma_{\ln{\Delta M},l,h}} \exp\left(\frac{\ln(\Delta M/\bar{\Delta M}_{l,h})^2}{2\sigma_{\ln{\Delta M},l,h}^2}\right), 
\end{equation}
with  $\bar{\Delta M}_{l}=0.3 \msun$,  $\bar{\Delta M}_{h}=1 \msun$, and $\sigma_{\ln{\Delta M},l}=\sigma_{\ln{\Delta M},h}=0.5$. 
Conservation of linear momentum implies that the velocity acquired by the CoM is:
\begin{eqnarray}\label{eq:CM_SN}
	\mathbf{V}_{\rm CM} &=& \mathbf{V}_{\rm CM,1}+\dfrac{m_2}{m_1+m_2} \mathbf{V}_{\rm kick,2}\nonumber\\
	&&+ \left(\dfrac{\Delta M}{m_1+m_2}\right) \left(\dfrac{m_1}{m_1+(m_2+\Delta M)} \mathbf{V}_{\rm kep}\right),
\end{eqnarray} 
where the second and third terms on the right hand side of the equation correspond to the contributions to the NSB CoM velocity from the kick experienced by the second remnant and from the (assumed instantaneous) mass change of the binary system \citep{Postnov2014}. 

The last parameter we estimate is the eccentricity $e$ of the NSB immediately after the second SN. $e$ is obtained assuming the conservation of energy and angular momentum after the second SN:
\begin{align}
	\dfrac{a_0}{a} &=  2 - \chi \left( \dfrac{V_{{\rm kick,2}, x}^2+V_{{\rm kick,2},z}^2+(V_{\rm kep}+V_{{\rm kick,2}, y})^2}{V_{\rm kep}^2}\right),\\
	1-e^2 &= \chi \left(\dfrac{a_0}{a}\right) \left(\dfrac{V_{{\rm kick,2},z}^2+(V_{\rm kep}+V_{{\rm kick,2},y})^2}{V_{\rm kep}^2}\right),
\end{align}
where $\chi = ((m_2+\Delta M) + m_1)/(m_2 + m_1) \geq 1$ is the fractional change in mass, and $V_{\rm kep} = \sqrt{G ((m_2+\Delta M)+m_1)/a_0}$.\footnote{Note that the assumption here on the direction of the secondary velocity immediately before the second SN is only made for clarity reasons. In the actual sampling $\mathbf{V}_{\rm kep}$ is assumed isotropic.}.

The time to coalescence due to GW emission $T_{\rm gw}$ is then evaluated from $(a,e,m_1,m_2)$ by performing the integration \citep{Peters_1964}
\begin{equation}
\label{eq:Tgw}
    T_{\rm gw}(a,e,m_1,m_2) = \dfrac{12 c_0^4}{19 \beta} \ \mathcal{G}(e)
\end{equation}
where
\begin{align}
    \beta &= \dfrac{64 G^3 m_1 m_2 (m_1+m_2)}{5 c^5},\nonumber\\
    c_0 &= \dfrac{a (1-e^2)}{e^{12/19}}\left(1+\dfrac{121}{304}e^2\right)^{-870/2299},\nonumber\\
    \mathcal{G}(e) &= \int_{0}^{e} {\rm d}e' \ \dfrac{e'^{29/19} \left( 1+121/304 e'^2 \right)^{1181/2299}}{(1-e'^2)^{3/2}}.
\end{align}
\begin{figure}
    \centering
    \includegraphics[scale=0.34]{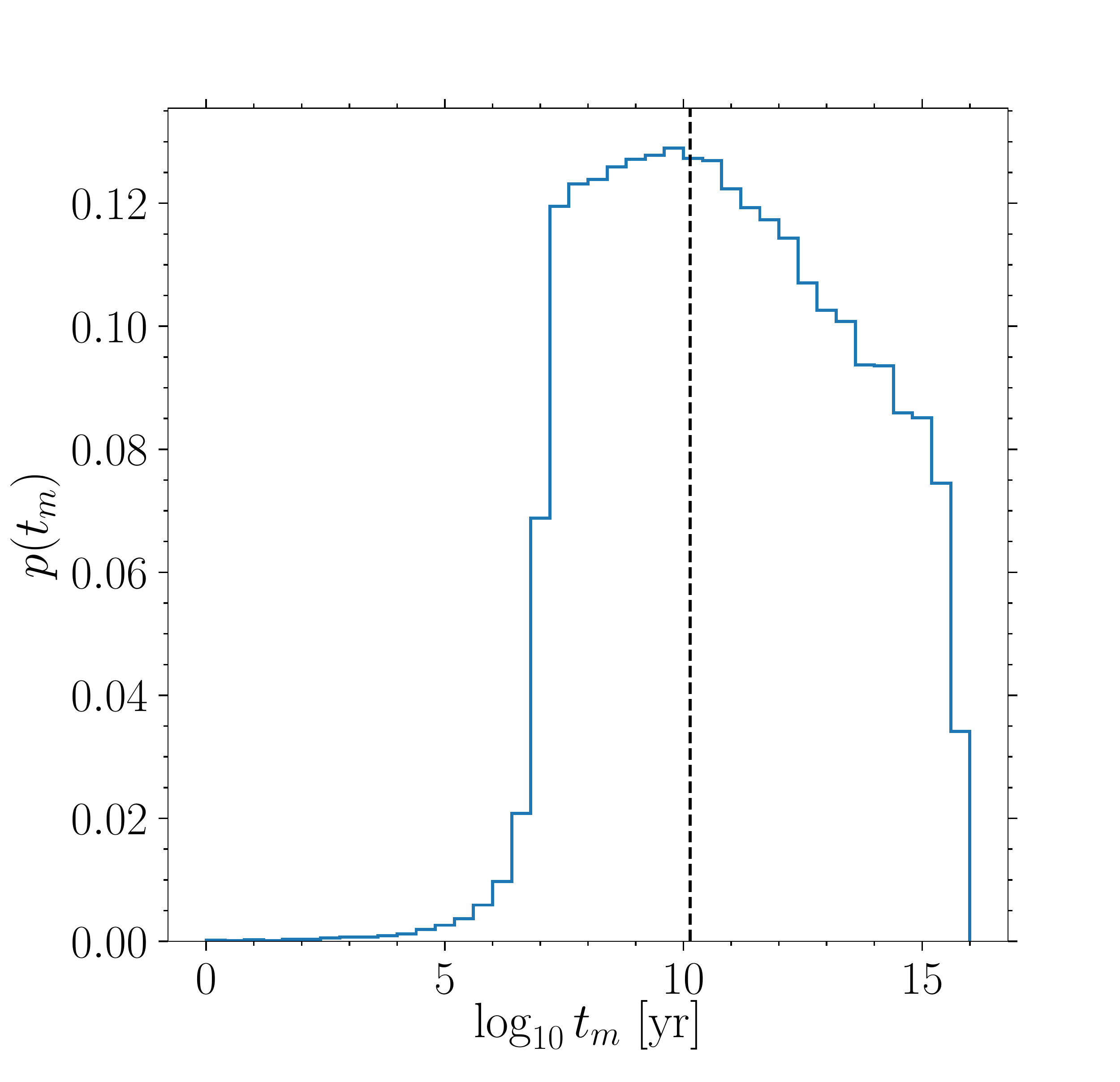}
    \caption{Probability density function of coalescence time for binaries generated in our model. Vertical dashed line indicates the current age of the universe. Despite the log-uniform distribution of the semi-major axis $p(t_m)$ is not flat because we do not assume zero eccentricity, determining a relatively fast coalescence also for system with larger initial semi-major axis. Note also that the lower limit adopted for the semi-major axis distribution determines the existence of fraction of rapidly-coalescing NSB systems.}
    \label{fig:Tmerger}
\end{figure}
In Fig.~\ref{fig:Tmerger} we present the probability distribution of the merger times, $t_{\rm m}$ computed as equation~\ref{eq:Tgw}, for a large set ($10^4$) of NSBs whose properties have been generated using the above mentioned distributions for the internal parameters and kick velocities. In particular, we consider the case $p_{\rm l} = 0.7$. In 42\% of the cases the NSB merges within the Hubble time. This percentage decreases to 27\% and 14\% if we limit $t_{\rm m}$ to 1~Gyr and 0.1~Gyr, respectively. We thus notice that our choice of a log-uniform distribution in the binary semi-major axis down to $a \lesssim R_{\odot} $ and the inclusion of highly eccentric cases (mainly for the high kick population) introduce a significant population of fast merging binaries \citep[see e.g.][]{Belczynski.etal:2002,OShaughnessy.etal:2008,Beniamini2016a}. However, this percentage is possibly smaller (by a factor $\lesssim 2$) than the one implied by the observed distribution of Galactic NSBs \citep{Beniamini.Piran:2019}.

\subsection{Binary NS external parameters and dynamical evolution}
\label{sec:external}

We sample the position of the CoM of the NSB at the time of the second SN explosion from the stellar disk profile (i.e. the exponential disk for dwarf galaxies or the superposition of MN for MW case), and we assume an initial CoM velocity 
\begin{equation}
    \mathbf{V}_{\rm CM,i} = \mathbf{V}_{\rm CM} + \mathbf{V}_{c},
\end{equation}
where $\mathbf{V}_{\rm CM}$ is the velocity acquired due to the SN kicks (see equation~\ref{eq:CM_SN}), while $V_{c}=\sqrt{R \,{\rm d}\Phi/{\rm d}r}$ is the circular velocity in the galactic plane (i.e. at $z = 0$) at the radius at which the NSB is initialized. 
We then integrate the motion of the CoM of the NSB in the chosen galactic system by numerically solving the equations of motion with an 8th order adaptive stepsize Dormand-Prince-Runge-Kutta algorithm (DOPRI853). We stop the integration at $T_{\rm gw}$ and we denote the NSB position at that time, $\mathbf{x}_f$, as the merger location. As a final step, we sample the direction of the NSB angular momentum from an isotropic distribution in order to infer the orientation of the binary orbital plane with respect to the galactic reference frame. 

\subsection{Binary NS ejecta: initial conditions and nucleosynthesis yields}
\label{sec:ejecta1}

The NSB is assumed to eject a certain amount of mass during and immediately after the merger. We consider two kinds of ejecta: dynamical and disc wind ejecta. 

\subsubsection{Dynamical ejecta}
Dynamical ejecta is expelled on a timescale of a few milliseconds due to tidal torques and hydrodynamics shocks. To model the properties of the dynamical ejecta, we use the parametric fit reported in \citet{Radice2018} for the total ejecta mass and average speed. To compute the NS compactness parameters associated with the two NS masses $m_{1,2}$ and required by these fits, we choose a nuclear equation of state compatible with all present nuclear and astrophysical constraints \citep[SLy,][]{Haensel2004}. Numerical simulations of NSB mergers show that this mass ejection happens preferentially along the equatorial plane. We thus adopt a $\sin^2{\theta}$ dependence of the mass spatial distribution on the polar angle measured with respect to the NSB rotational axis \citep{Perego2017a,Radice2018}. Moreover, neutrino irradiation, more intense inside the polar funnels, increases the electron fraction above 0.25 for $\theta \lesssim \pi/4$ and $\theta \gtrsim 3\pi/4$. As a consequence, we assume that polar ejecta produces $r$-process nucleosynthesis yields between the first and second $r$-process peaks, for which  we consider an atomic mass number interval $70 \leq A \leq 120$. On the other hand, the more neutron rich equatorial ejecta produces elements between the second and third $r$-process peaks for which we assume an $r$-process nucleosynthesis interval $120 \leq A \leq 232 $.

\subsubsection{Disc wind ejecta}

The ejection of matter in the form of disc winds happens on timescales longer than the ones of the dynamical ejecta (up to a few hundreds milliseconds). It is due to neutrino absorption, magnetic and viscous processes inside the remnant.
To estimate the total mass contained inside this ejecta, we assume that a fixed fraction of the disc $\xi_{\rm wind}$ becomes unbound and is launched with an average velocity $v_{\rm disk}$. In this work we assume $\xi_{\rm wind} = 0.2$ and $v_{\rm disk} = 0.08 {\rm c}$ \citep[see e.g.][]{Just2015,Fernandez2018}.
The disc mass is determined through the fitting formula reported in \citet{Radice2018}. 
As in the case of the dynamical ejecta, the calculation of the dimensionless tidal coefficient of the merging NSB required by this fitting formula is computing assuming the same SLy nuclear equation of state used for the dynamical ejecta.
This ejecta is expected to be more isotropic, both in terms of mass and electron fraction distribution. However, neutrinos could also affect its nucleosynthesis \citep{Perego2014b,Martin2015,Lippuner2017}.
In particular, if the merger results in a long-lived massive NS, the ejecta electron fraction could be systematically shifted above 0.25 such that the production of heavy $r$-process elements is prevented and the nucleosynthesis produces only nuclei between the first and second $r$-process peaks, i.e. for $ 70 \leq A \leq 120$. Otherwise, if a black hole forms promptly or on a timescale smaller than the disc viscous timescale, the production of all $r$-process elements is foreseen. In this case we assume a mass number interval $ 80 \leq A \leq 232$ for the $r$-process nucleosynthesis. To distinguish between the first and the second case we use an empirical threshold value suggested by the CoRe NSB merger database \citep{Dietrich2018}: if $(m_1 + m_2) < 1.3 M_{\rm NS,max}$ (where $M_{\rm NS,max} \approx 2.05 \msun$ is the maximum cold NS mass predicted by the SLy nuclear equation of state) then the central remnant does not collapses to a BH before the wind ejecta is expelled and the production of all $r$-process elements in the disk ejecta is prevented.

\subsection{Binary NS ejecta: evolution}
\label{sec:ejecta2}

The ejecta expanding in the intergalactic (IGM) or interstellar (ISM) medium will slow down, forming a NSB merger remnant similar to the remnant produced by SN explosions \citep[see e.g.][]{Montes2016}. This will eventually mix with the surrounding medium, enriching it with its nucleosynthesis yields. We assume the IGM/ISM to follow the same density profile as the model galaxy down to a floor number density $n_0$, and to be formed by hydrogen atoms with a temperature of $T \sim 10^4 {\rm K} $.
The relation between the mass and the number floor densities is given by $\rho_0 = n_0 m_p \mu_a $, where $m_p$ is the proton mass and $\mu_a$ the mean atomic weight, $\mu_a = 1.27$. Values of $n_0$ around galaxies are largely unknown and strongly dependent on the environment and cosmological epoch.
We consider a fiducial value $ n_0 = 10^{-4} {\rm cm^{-3}}$, larger but still comparable to the present average density of the Universe ($\sim 2\times 10^{5}{\rm cm^{-3}}$). To explore the possible impact of larger densities, expected for example in the early Universe, we investigate also $ n_0 = 10^{-2} {\rm cm^{-3}}$. We notice that the resulting interval 
$ 10^{-4} {\rm cm^{-3}} \lesssim n_0 \lesssim 10^{-2} {\rm cm^{-3}}$ is compatible with the values inferred by the GRB afterglow emission of GW170817 \citep{Margutti2017,Ghirlanda2019}.

To model the evolution of the remnant, we assume that the two kinds of ejecta (characterized by different initial kinetic energies) will produce two remnants, that we treat independently. Moreover, for simplicity, we consider the remnant expansion to happen inside a uniform medium of density 

\begin{equation}
    \rho_{\rm med} = \mathrm{max}(\rho_d(\mathbf{x}_f), \ n_0 m_p \mu_a),
\end{equation}
where $\mathbf{x}_f = (R_f,z_f)$ represents the merger location in the galactic reference frame with $\rho_d(\mathbf{x}_f)$ denoting the baryonic density (cf. section~\ref{sec:dwarf}) at such point.
If $R_{\rm rem}$ and $v_{\rm rem}$ denote the radius and the speed, respectively, of the NSB remnant forward front, we model their time evolution as in the case of SN remnants and we follow an approach similar to the one described in \citet{Haid2016} and in \citet{Beniamini2018}.
More specifically, we consider the following expansion phases:
\begin{itemize}
    \item free expansion: during this phase, the ejecta expands with constant velocity $v_{\rm ej}$ equal to either $v_{\rm dyn}$ or $v_{\rm disk}$, depending on the nature of the ejecta considered:
    \begin{equation}
    \begin{cases}
        R_{\rm rem}(t) = v_{\rm ej} t \quad 0 \leq t \leq t_1 \, , \\
        v_{\rm rem}(t) = v_{\rm ej}   \quad 0 \leq t \leq t_1 \, .
    \end{cases}
    \end{equation}
    This phase lasts up to $t_1 = \frac{1}{v_{\rm ej}} \left( \frac{3 m_{\rm ej} }{4 \pi \rho_0} \right)^{1/3}$, i.e. the point when the ejecta has swept-up a mass comparable to its total mass. 
    \item Sedov-Taylor expansion: this phase lasts up to the point where radiative cooling becomes relevant. This transition usually happens at $t \sim t_{\rm TR} = 4 \times 10^4 \, {\rm yr} \, n_0^{-0.53}$ \citep{Haid2016}. The phase $ t_{\rm TR} \leq  t < C_{\rm TR} t_{\rm TR} $, with $C_{\rm TR} \approx 1.83$, is an intermediate phase in which the expansion cannot be expressed as a self-similar solution. In this first study we neglect this additional complication and assume that the self-similar Sedov-Taylor expansion continues up to $t_2 = C_{\rm TR} t_{\rm TR}$
    \begin{equation}
    \begin{cases}
        R_{\rm rem}(t) = R_{\rm rem}(t_1) + \xi \left( \frac{E_{\rm rem}}{\rho_0} \left( t-t_1 \right)^2 \right)^{1/5}  \quad t_1 < t \leq t_2 \, ,\\
        v_{\rm rem}(t) = \frac{2}{5} \xi \left( \frac{E_{\rm SN}}{\rho_0} \right)^{1/5} \left( t-t_1 \right)^{-3/5}   \quad t_1 < t \leq t_2 \, ,
    \end{cases}
    \end{equation}
    where $\xi = 25/(4 \pi)$ and $E_{\rm rem} \approx m_{\rm ej}v_{\rm ej}^2/2$ is the initial kinetic energy of the ejecta.
    \item snowplow expansion: in this phase, the ejecta has produced a thin shell, containing most of the ejecta mass, that expands driven 
    by the hot interior (pressure-driven snowplow phase). This phase is characterized again by a self-similar solution, and continues up to 
    the point where the ejecta velocity equals the sound speed of the IGM/ISM, $v_{\rm lim} \approx 10~{\rm km/s}$.
    If we denote this time as, 
    \begin{equation}
        t_3 = \left( \frac{2}{7 v_{\rm lim}} \frac{R_{\rm rem}(t_2)}{t_2^{2/7}} \right)^{7/5},
    \end{equation}
    then
    \begin{equation}
    \begin{cases}
        R_{\rm rem}(t) = R_{\rm rem}(t_2) \left( \frac{t}{t_2} \right)^{2/7}  \quad t_2 < t \leq t_3 \, ,\\
        v_{\rm rem}(t) = \frac{2}{7} \frac{R_{\rm rem}(t_2)}{t_2} \left( \frac{t}{t_2} \right)^{-5/7}   \quad t_2 < t \leq t_3 \, .
    \end{cases}
    \end{equation}  
    The maximum remnant expansion is assumed to occur at $t_3$, since after that the ejecta dissolves into the IGM/ISM.
\end{itemize}


\begin{figure}
    \centering
    \includegraphics[width=0.47\textwidth]{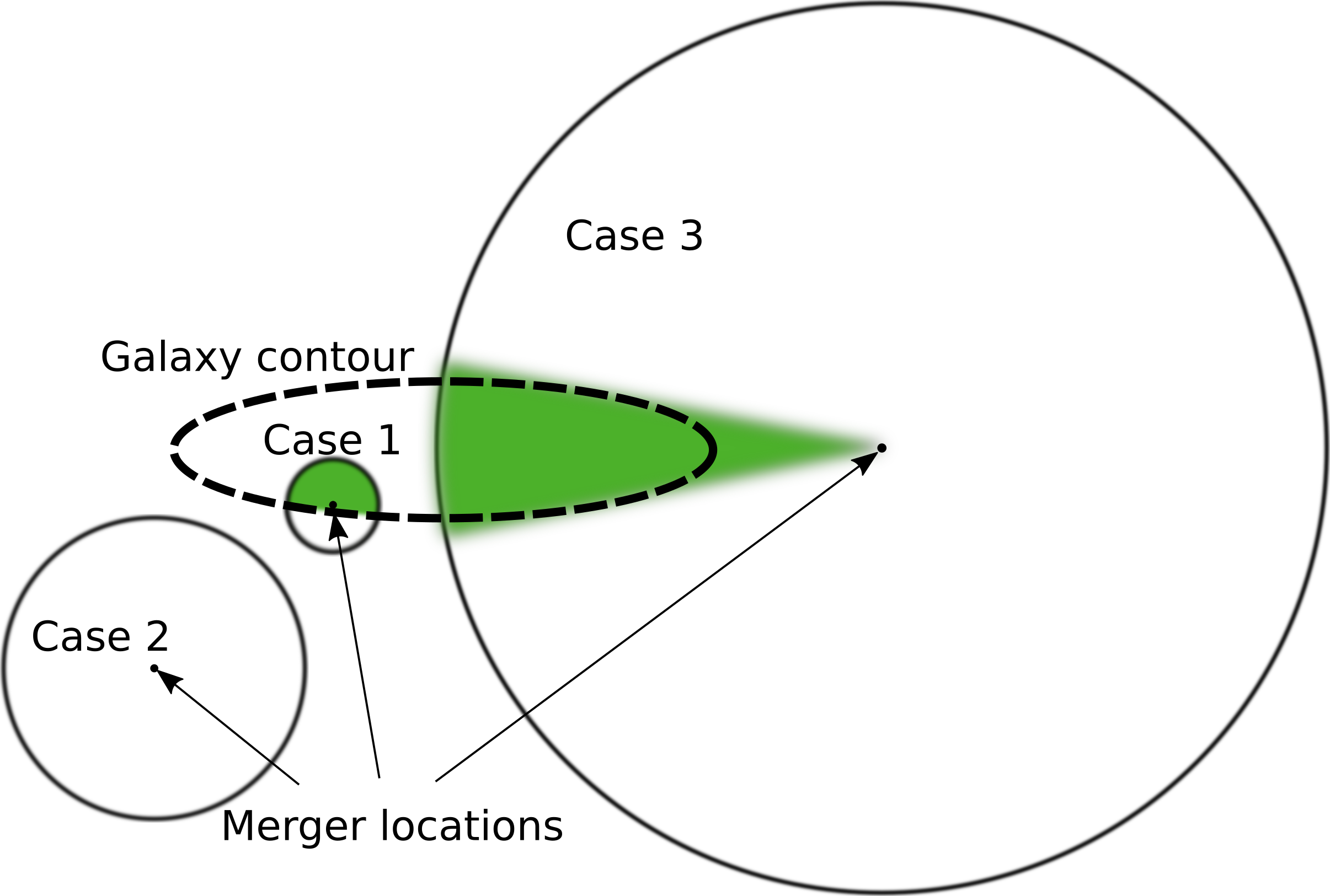}
    \caption{Cartoon representation of the three possible situation that determine the amount of $r-$process enrichment (see section~\ref{sec:ret_mass}).}
    \label{fig:cartoon}
\end{figure}

\section{Retained ejecta and elemental abundances}
\label{sec:retained ejecta}

\subsection{Fraction of retained ejecta}
\label{sec:ret_mass}

Let us consider a NSB that merges at a certain location $\mathbf{x}_f = (x_f,y_f,z_f) = (R_f,z_f)$. 
If the resulting remnant is large enough (or alternatively the merger point is not too far from the host galaxy), a certain fraction of the ejecta can
be retained by the galaxy. Moreover, if the NSB inspiral and the remnant evolution are fast enough (i.e. $T_{\rm GW} + t_{3} \lesssim T_{\rm SF}$), the retained mass will pollute the gas that will form a new generation of stars. To compute the fraction of retained ejecta, the first step is to define a suitable contour for the galaxy. We define it as the iso-density contour in the baryonic density profile $\rho_{\rm lim} = \rho_d ( \xi_{\rm lim} R_{\rm d},0 )$ and we consider the surface $\mathcal{S}$ (enclosing the volume $\mathcal{V}$, i.e. the galaxy) defined by all points $(R_p,z_p)$ that satisfy the condition $\rho_d ( R_p,z_p ) = \rho_{\rm lim}$. We set $\xi_{\rm lim} = 3$ and $\xi_{\rm lim} = 5$ in order to explore the impact of the chosen threshold on the retained mass.\footnote{Here we comment on the dwarf galaxy case only (i.e. $\rho_d$ is given by the exponential disk), with the understanding that the procedure followed for the MW is conceptually the same.}
If $\rho_{\rm med}$ denotes the density inside which the remnant expands (see previous subsection), these are the possible scenarios (see Fig.~\ref{fig:cartoon} for a cartoon sketch):
\begin{itemize}

\item if $\rho_d(\mathbf{x}_f) > \rho_{\rm lim}$, the coalescence happens within the host. When the ejecta reaches the maximum expansion radius, $R_{\rm rem}(t_3)$, we evaluate the intersection of the corresponding spherical shell $\Sigma$ with the volume $\mathcal{V}$ and we estimate the amount of the retained mass as the fraction of the spherical surface that lies inside the galaxy times the total ejected mass, i.e.
\begin{equation}
    m_{\rm ret} = \textrm{max}\left( \int_{\Sigma \cap \mathcal{V}} \left( \frac{{\rm d}m_{\rm ej}}{{\rm d}\Sigma} \right) {\rm d}\Sigma \ ,0.5 m_{\rm ej, tot}\right),
\end{equation}
where ${\rm d}m_{\rm ej}/{\rm d}\Sigma$ is the ejecta angular distribution on the
remnant sphere.
The maximum with $0.5 m_{\rm ej, tot}$ ensures that at least half of the ejected mass remains inside the galaxy. This is due to the fact that the increasing density gradient moving from $\mathbf{x}_f$ towards the galactic disc plane will certainly slow down the ejecta moving in that direction. 

\item if $\rho_d(\mathbf{x}_f) < \rho_{\rm lim}$ and $\Sigma$ does not expand across the galaxy volume $\mathcal{V}$, then none of the ejecta can enrich the galaxy. Practically, the situation of non-intersection verifies when one of conditions below is satisfied
 \begin{align}
    \sqrt{x_f^2+y_f^2} - R_{\rm rem}(t_3) & > \xi_{\rm lim}R_d, \nonumber\\
    |z_f - R_{\rm rem}(t_3)| & > |z_{\rho_{\rm lim}}|,
 \end{align}
where $\xi_{\rm lim}R_d$ is the largest disk contained inside $\mathcal{V}$, while the interval $[-z_{\rho_{\rm lim}},z_{\rho_{\rm lim}}]$ represents the maximum $z$ extent of the galaxy at a radial distance $R_f=\sqrt{x_f^2+y_f^2}$. 
 
\item if $\rho_d(\mathbf{x}_f) < \rho_{\rm lim}$ and the remnant sphere extends inside or beyond the galaxy, then a certain amount of the ejected mass can pollute the galaxy. In order to estimate this mass, we identify the fraction of solid angle that intersects the galaxy, $\Omega_f$, using a Monte Carlo approach.\footnote{To this end, we generate $\sim 4 \times 10^{6}$ propagation directions starting from $\mathbf{x}_f$ and spanning the whole solid angle.}
The retained mass is then computed as 
\begin{equation}
    m_{\rm ret} = \int_{\Omega_f} \left( \frac{{\rm d}m_{\rm ej}}{{\rm d}\Omega} \right) {\rm d}\Omega \,,
\end{equation}
where ${\rm d}m_{\rm ej}/{\rm d}\Omega$ is the ejecta angular distribution.

\end{itemize}

\subsection{\textit{r}-process nucleosynthesis abundances}

As stated in section \ref{sec:ejecta1}, NSB merger ejecta is characterized by a heterogeneous composition of heavy $r$-process elements potentially produced in different components. In order to asses the contribution of each element to the mass retained by the host galaxy, we consider that the relative abundances of heavy elements closely follows the Solar System (SS) abundances obtained from \citet{Lodders2003} and decomposed in its $s-$ and $r-$process contributions following \citet{Sneden2008}. If $A_{\rm min}$ and $A_{\rm max}$ are the minimum and maximum mass number produced by the $r$-process nucleosynthesis in a certain ejecta component, the mass fraction of each element E$_i$ (whose isotopes are such that $A_{\rm min} \leq A_i \leq A_{\rm max}$) inside the retained mass of each component can be expressed as
\begin{equation}
	X(E_i) = \frac{m_{\rm ej,ret}(E_i)}{m_{\rm ej,ret}} = \dfrac{ \left( Y(E_i) \right)_{\odot} \langle A_i \rangle}{\sum_{k} \left( Y(E_k) \right)_{\odot} \langle A_k \rangle },
\end{equation}
where $Y({E_i})$ is the elemental abundance of E$_{i}$, $\langle A_i \rangle$ its mean mass number, as obtained by the SS abundances, and the sum at denominator ranges over all produced elements.
\footnote{Throughout the literature, the SS abundance of an element E$_i$ is usually quoted in terms of the astronomical logarithmic scale $A(\textrm{E}_i) = \log_{10}(n_i/n_H) + 12$, with $n_H$ the number density of Hydrogen atoms or according to the cosmochemical scale, which instead normalises the number of Silica atoms to $10^6$. The relation between the two scale is given by $A(\textrm{E}_i) = 1.54 + \log_{10} N_{\rm cosmo}(E_i)$.}
Once $A_{\rm min}$ and $A_{\rm max}$ are given, $X(E_i)$ is unambiguously determined and allows us to obtain the mass composition of the $r$-process nucleosynthesis retained material in each component.

The total retained mass of each element is then obtained as the sum of the element retained masses in all components. Finally, the abundance ratio of two elements, defined as 
\begin{equation}
    [{\rm E_i/E_j}] = \log_{10}\left(\dfrac{Y(E_i)}{Y(E_j)}\right) - \log_{10}\left(\dfrac{Y(E_i)}{Y(E_j)}\right)_\odot,
\end{equation}
can be equivalently computed as
\begin{equation}
    [{\rm E_i/E_j}] = 
    \log_{10} \left( \dfrac{m(E_i) \langle A(E_j) \rangle}{m(E_j) \langle A(E_i) \rangle } \right) 
    - \log_{10}\left(\dfrac{n_i}{n_j}\right)_\odot,
\end{equation}
where the last term is a constant determined by the SS abundances.

In addition to the retained mass in $r$-process elements, we also compute the iron mass produced by CCSNe. According to observations, 2/3 of CCSNe explode as type II SN and produce, on average, 0.02~$\msun$ of iron; 1/3 are SNIbc and produce, on average, 0.2~$\msun$ of iron \citep{Li2011,Drout2011}. For each galaxy realization, if $N_{\rm CCSN}$ is the number of CCSNe occurred within a time $t$, we sample $N_{\rm CCSNII}$ type II SNe from a Poisson distribution with average equal to $2 N_{\rm CCSN}/3$, and compute $N_{\rm CCSNIbc} = N_{\rm CCSN}-N_{\rm CCSNII}$. Since CCSNe are expected to explode inside the galaxy, we further assume that all the iron is retained inside the galaxy and we estimate its amount from the average quantity produced per event.
More detailed studies taking into account the kinetic energy of the ejecta as well as the elemental mixing \citep{Beniamini2018,Emerick2018}, showed that the amount of retained elements could sensitively depend on the source. Thus, our values provide upper limits to the iron enrichment due to CCSNe.
For example, \citet{Beniamini2018} estimated a retain factor of 0.2-0.9 for ejecta products exploded within a dwarf galaxy.

\section{Results}
\label{sec:results}

\begin{figure*}
    \centering
    \includegraphics[width=\textwidth]{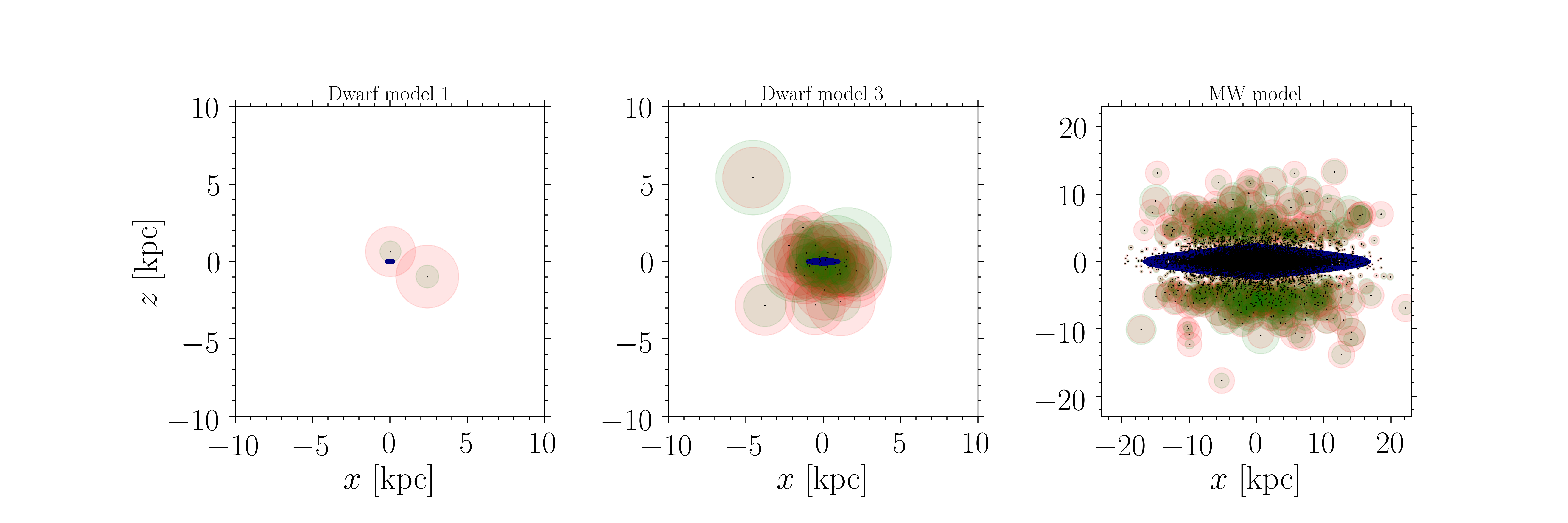}
    \caption{Projection in the $x-z$ galactic frame of NSB coalescence positions (black dots) and expansion sphere of ejected material. Dynamical ejecta are displayed in red, while wind ejecta in green. {\it Left panel}: Dwarf 1 case. {\it Central panel}: Dwarf 2 case. {\it Right panel}: MW-like case. In all panels the dark blue profiles represent the galaxy density iso-contour equal to the chosen $\rho_{\rm lim}$ (see section~\ref{sec:ret_mass} for details).}
    \label{fig:bubble}
\end{figure*}

In this section we present the result obtained when all the ingredients presented in sections~\ref{sec:method}-\ref{sec:retained ejecta} are combined. 
We first focus on a fiducial case characterised by $n_0 = 10^{-4}{\rm cm^{-3}}, \xi_{\rm lim} = 3, x = 150, p_l = 0.7$ in Sec.~\ref{sec:results, model exploration}. We then discuss the impact of parameters by varying each of them one at a time in Sec.\ref{sec:parameter_exploration}. 

\begin{figure*}
    \centering
    \includegraphics[width=0.8\textwidth]{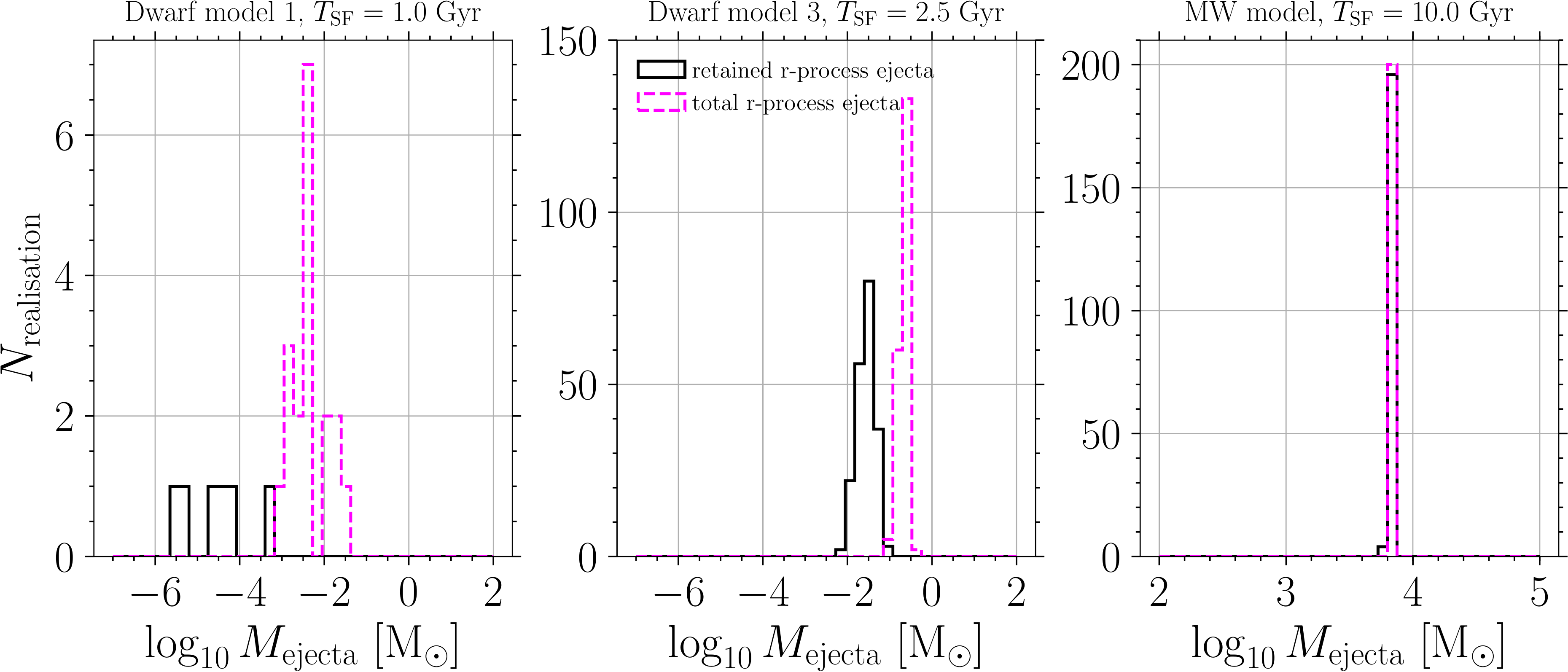}
    \caption{Comparison between distributions of the total ejected mass (dashed purple lines) compared to the actual retained mass (solid black line) for the standard case. {\it Left panel}: Dwarf model 1. {\it Central panel}: Dwarf model 3. {\it Right panel}: MW model. Note how the discrepancy between the two masses decreases with increasing galaxy mass.}
    \label{fig:ejecta_vs_ret}
\end{figure*}

\subsection{Model exploration}
\label{sec:results, model exploration}

A first qualitative interpretation of our results is given by Fig.~\ref{fig:bubble}, where we report the $x-z$ projection of the NSB merger locations (black dots in the figure) as well as the expansion sphere (circles in the projections) of the ejected $r-$process material. In this plot and for the other that follow, we consider, from left to right, three galaxy models with total baryonic mass of $10^5,10^7,\sim 5\times 10^{10}\msun$, indicatively representative of UFDs, classical dwarf and MW-like galaxies. From the figure it is evident that two factors primary determine the level of enrichment of the progenitor galaxy (dark blue contour in the figure): the number of NSBs that form in the galaxy  and merge within $T_{\rm SF}$, and the distance from the galaxy at which each NSB travels prior to coalescence. Both factors crucially depend on the baryonic mass of the galaxy. In very low mass galaxies (e.g. left panel of Fig.~\ref{fig:bubble}) the enrichment level is quite low since the average number of formed NSB is usually small, around unity or fractions of it. In addition, the potential well provided by the galaxy is rather shallow, therefore the SN kicks can imprint enough velocity such that a NSB system can substantially recede from the parent galaxy. Even if a NSB is gravitationally bound to the galaxy, it often spends most of its orbital time rather far from the galactic disc. On the contrary, for massive galaxies (e.g. MW-like ones, right panel of Fig.~\ref{fig:bubble}) the level of enrichment is much higher given the increased number of massive stars that can produce NSB systems, the longer $T_{\rm SF}$, as well as the higher escape velocity that prevent NSB to travel far away from the galaxy (the energetics of the SNae is reasonably assumed to be the same irrespective of the galaxy mass). Between the above situations, galaxies with intermediate masses continuously connect the two extremes, as visible for example in the central panel of Fig.~\ref{fig:bubble}. 

This can be more quantitatively inferred from Fig.~\ref{fig:ejecta_vs_ret}, in which we report the mass distribution of the whole amount of produced $r-$process material (dashed magenta lines) compared to the $r-$process mass actually retained by the host galaxy (solid black lines). For each galaxy we also consider only NSBs that coalesce within the time available for star formation ($T_{\rm SF}$), since a galaxy with no more star formation activity cannot form stars enriched with heavy elements. 

\begin{figure*}
    \centering
    \includegraphics[width=0.8\textwidth]{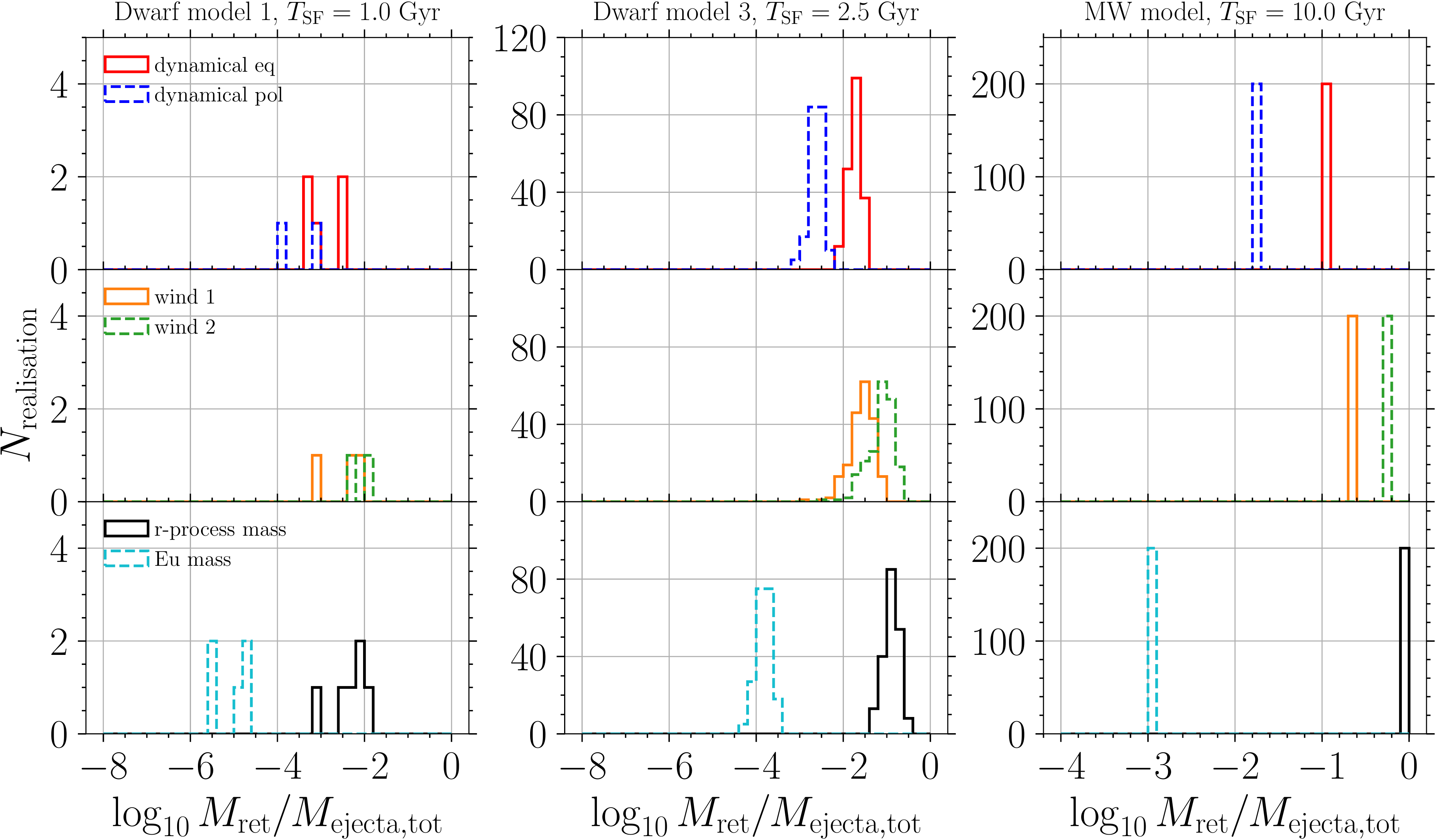}
    \caption{Distribution of retained mass that enriches the galaxy over total ejected mass for a CCSN to NSB ratio $x=150$. {\it Left panels}: Dwarf model 1. {\it Central panels}: Dwarf model 3. {\it Right panels}: MW model. {\it Top panels}: dynamical polar ejecta (blue dashed line) and dynamical equatorial ejecta (red solid line). {\it Central panels}: wind 1 ejecta (green dashed line) and wind 2 ejecta (orange solid line). {\it Bottom panels}: total retained $r-$process (black solid line) and total retained Europium (cyan dashed line).}
    \label{fig:standard_case}
\end{figure*}

In the least massive galaxy case ($M_{\rm b} = 10^{5}\msun$, left panel), only a small fraction of the realizations ($< 10\%$) presents at least one NSB merger happening within $T_{\rm SF} = 1~{\rm Gyr}$. The small probability that the corresponding ejecta intersects the galactic disc further reduces the probability that a galaxy realization displays a $r$-process material enrichment ($\sim 2\%$).
For more massive galaxies the distributions are more peaked with an higher number of realisations providing some enrichment, meaning that in these systems NSB mergers are more likely and generally closer to the galaxy.
Moreover, from the figure it is particularly clear that there exists a dilution factor between the $r-$process mass produced by NSB mergers and the actual amount captured by the host galaxy. This dilution factor is simply due to the weakening of the galactic gravitational field and increases with decreasing galaxy mass. For instance, in the least massive galaxy case the amount of retained material is diminished by at least a factor 10 with respect to the produced one. Overall, there is a potential discrepancy of more than three order of magnitude for the lightest galaxies between the produced and the retained $r$-process material. For larger galaxy masses, as of a consequence of the larger numbers of merging NSBs and of the deeper gravitational potential of the galaxy, the amount of $r$-process material increases and the discrepancy between the produced and the retained material decreases. A global dilution factor of the order of ten is still visible for the $M_{\rm b}=10^7 \msun$ model (central panel). Finally, no significant discrepancy is visible for the most massive case (right panel), where a large number of NSB merger occur within $t = 10~{\rm Gyr} < T_{\rm SF}$, most of them pollute the galaxy, and no significant dilution factor applies. In the following, unless differently specified, we will always refer to $r-$process mass actually retained by galaxies.

\begin{figure*}
    \centering
    \includegraphics[width=0.8\textwidth]{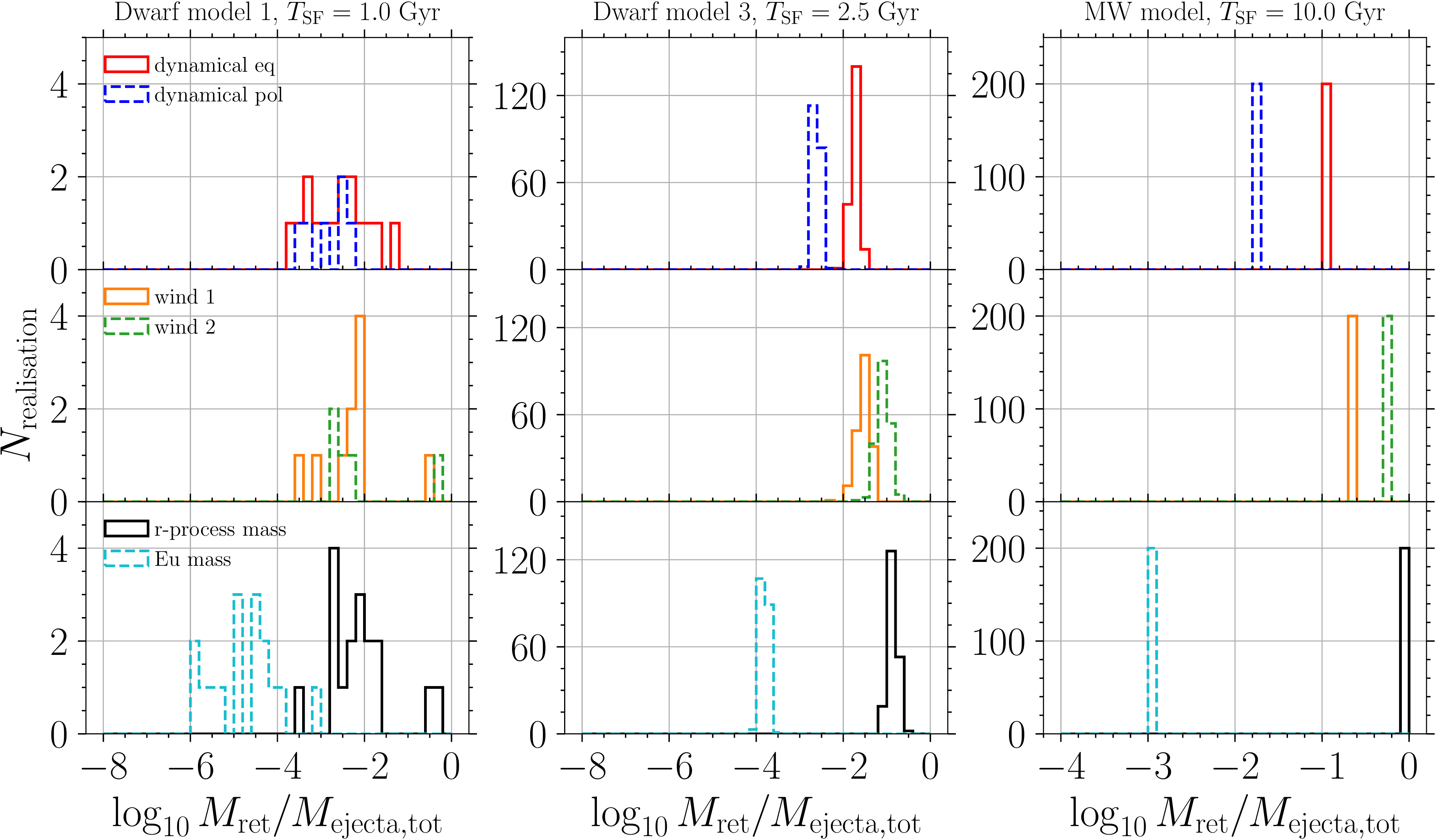}
    \caption{Same as Fig.~\ref{fig:standard_case}, but considering the optimistic case with $x=50$.}
    \label{fig:opt_case}
\end{figure*}
\begin{figure*}
    \centering
    \includegraphics[width=0.8\textwidth]{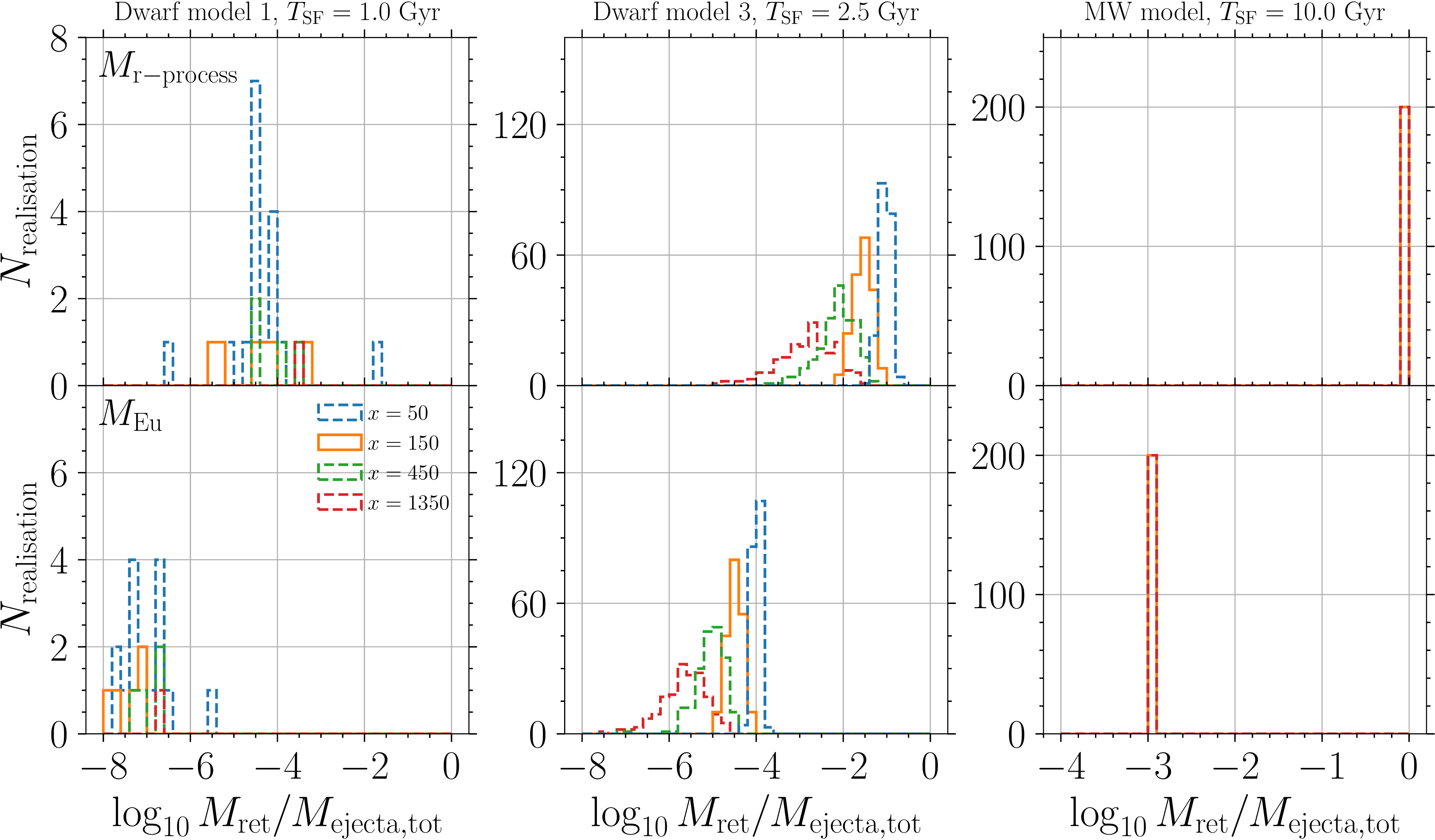}
    \caption{Distribution of retained $r-$process mass (upper panels) and retained europium mass (lower panels) over total ejected mass that pollute a selected galaxy model (as labelled) when considering a different ratio $x$ between the number of CCSN and the actually formed NSB. We explore a range from $x = 50$ (optimistic) to $x = 1350$ (extremely pessimistic).}
 \label{fig:var_X}
\end{figure*}

\begin{figure*}
    \centering
    \includegraphics[width=0.8\textwidth]{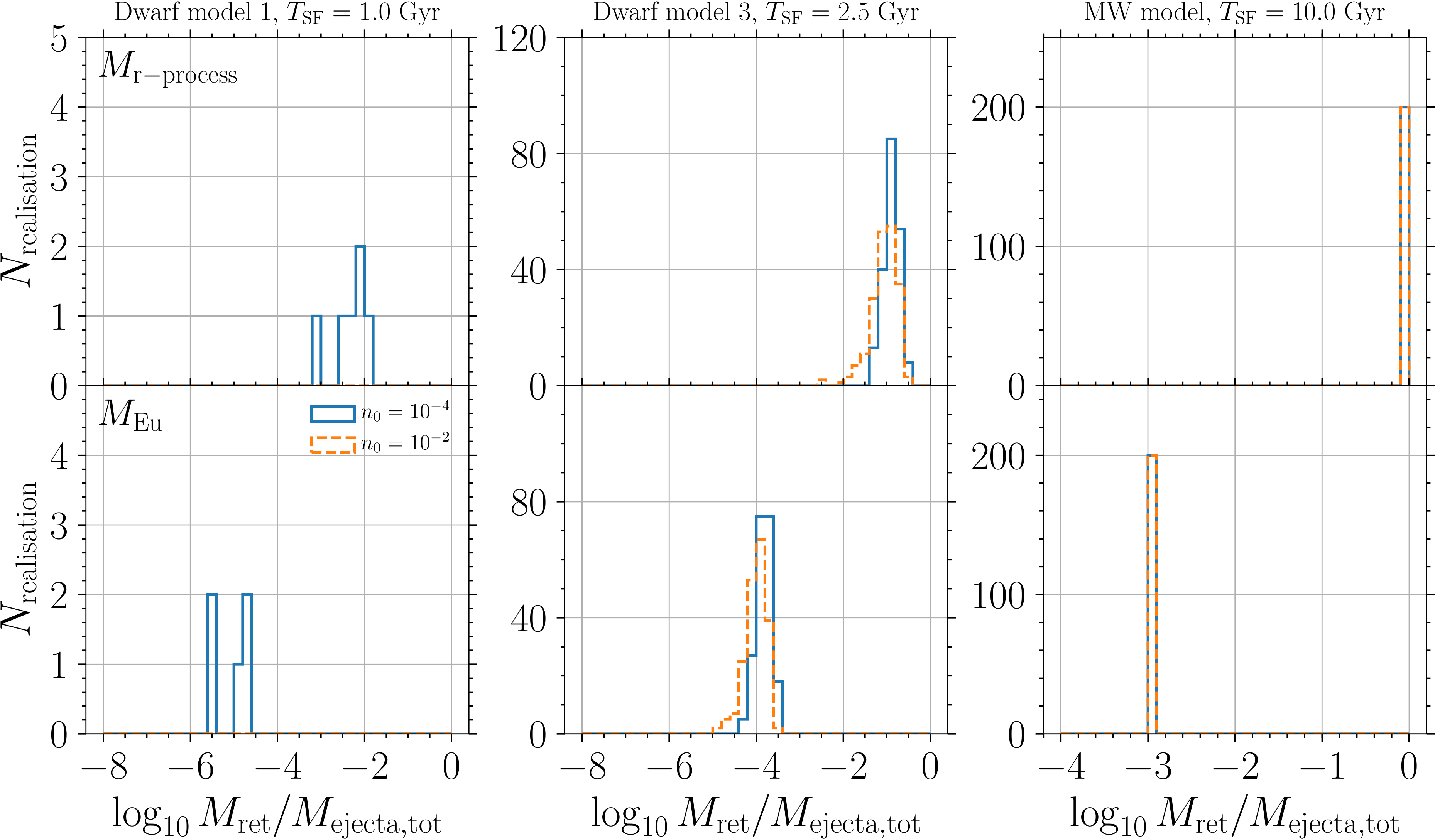}
    \caption{Same as Fig.~\ref{fig:var_X}, but considering a higher mean IGM density of $n_0 = 10^{-2}{\rm cm^{-3}}$, instead of $n_0 = 10^{-4}~{\rm cm^{-3}}$ used for the standard case. Note how an higher $n_0$ prevents any $r-$process enrichment in the least massive galaxy model, where generally NSB can cover large distances before merging.}
    \label{fig:var_n0}
\end{figure*}

A deeper insight into the specific origin of $r-$process material is given in Fig.~\ref{fig:standard_case} and Fig.~\ref{fig:opt_case}, for the CCSN to NSB ratio $x=$~150 and 50, respectively. Upper panels show $r-$process mass generated from dynamical ejecta, both from equatorial (solid red lines) and polar components (dashed blue lines). Middle panels report instead the mass coming from disk wind ejecta, labelled as ``wind 1'' for the wind coming from more massive NSB that do  form a BH before the disk wind emerges (solid orange lines) and ``wind 2'' for the ejecta coming from NSB merger that form a long-lived NS (dashed green lines). We recall that the latter cases do not produce a significant amount of Eu. Finally, lower panels show the distributions of total retained $r-$process mass (solid black lines) as well as retained europium mass (dashed cyan lines). 
For the dynamical ejecta, the $\sin^2\theta$ mass distribution, as well as the larger solid angle, favours pollution from the equatorial component, inside which europium is produced. For the disk winds, lighter NSB mergers produce more massive disc whose ejecta can pollute the galaxy more significantly, but without producing significant amount of heavy $r$-process elements, including europium. Summing up of the relevant contributions, the amount of retained europium is thus usually three orders of magnitudes smaller than the total amount of $r$-process material.
Results for the more optimistic case ($x=50$, Fig.~\ref{fig:opt_case}) are qualitatively very similar to the $x=150$ ones, but showing an higher number of successful events and larger amount of $r$-process material and europium masses. Still for the least massive case the ratio of retained vs produced europium mass remains usually very low, i.e. around $10^{-6}-10^{-5}$. 
Once again, larger galaxy masses reduce fluctuations in the distributions, leading to narrow histograms in the MW-like cases. 

\subsection{Parameter exploration}
\label{sec:parameter_exploration}

We now turn to explore the dependence of our results on some of the key parameters of the model.

\begin{figure*}
    \centering
    \includegraphics[width=0.8\textwidth]{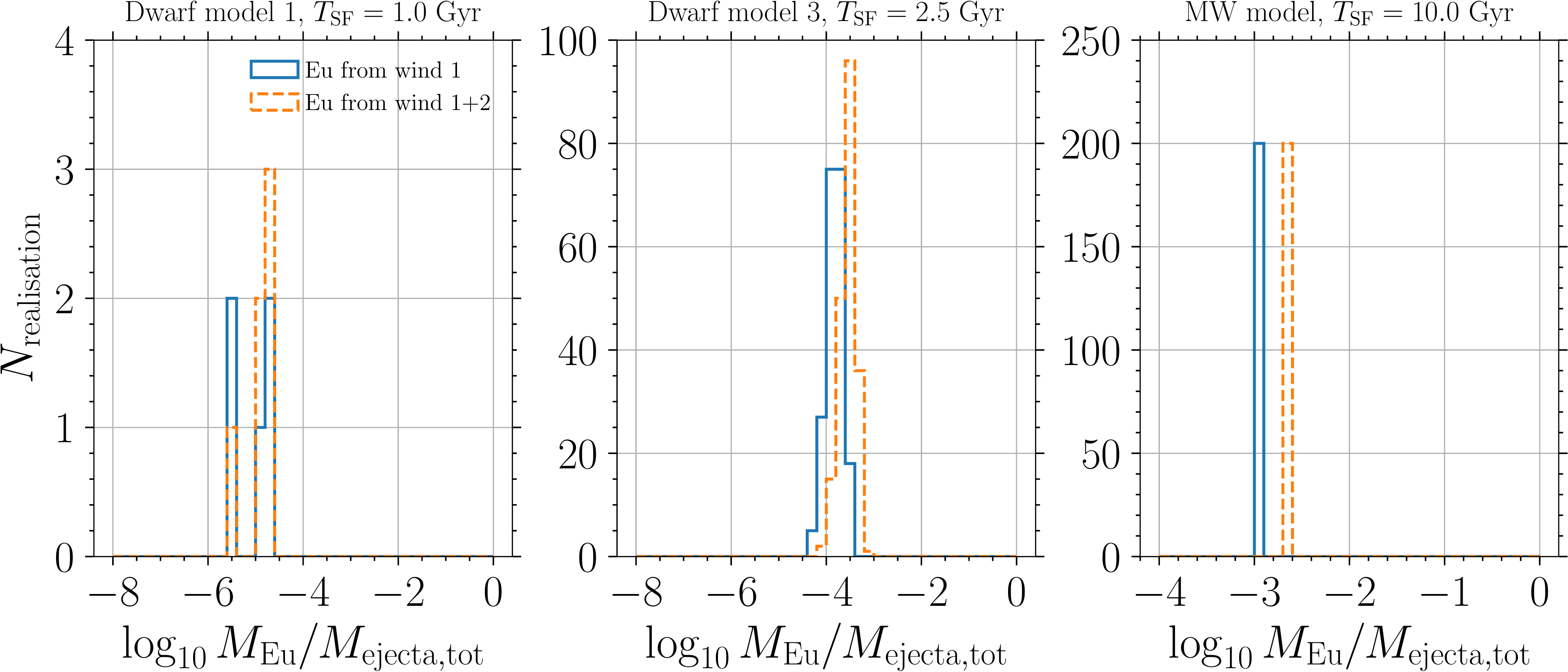}
    \caption{Distribution of retained europium mass over total ejected $r-$process mass when its production is allowed also from long-lived NS (wind 2 case).}
    \label{fig:var_wind}
\end{figure*}

In Fig.~\ref{fig:var_X} we present the distribution of $r$-process material for different $x$, i.e. the ratio between CCSNe and formed NSBs. Also for the least massive galaxy the increased number of potential NSB mergers leads to an increment of the retained mass, but still revelling a high scatter, meaning that the enrichment process is dominated by small number statistic. A clear trend is instead observable in the other two cases (central and right panels of Fig.~\ref{fig:var_X}) showing an increase of nearly two orders of magnitude in the peak of retained mass when going from $x=1350$ (very pessimistic) to $x=50$ (optimistic).

Another crucial parameter that can affect the amount of $r-$process enrichment, especially at low galactic masses, is the IGM density. If NSBs merge outside from the galaxy, a higher IGM density determines a smaller expansion of the ejecta bubble, preventing the ejected material to fall back on the parent galaxy. Such trend is shown in Fig.~\ref{fig:var_n0}, in which we compare our standard case with an IGM number density of $n_{0} = 10^{-4}~{\rm cm^{-3}}$ (solid blue lines) to a situation with $n_0$ one hundred times higher. No sensible effect is seen in the MW case, while for Dwarf model 3 we witness a slightly decrease in the number of successful enrichment realisations and, at the same time, a decrease in the efficiency of the $r$-process enrichment. Very different is instead the case of Dwarf model 1 where assuming $n_0 = 10^{-2}~{\rm cm^{-3}}$ no $r-$process enrichment is verified (with 200 galaxy realisations).

In addition, for the europium mass only, we verify the dependence on our assumptions about the the $r-$process nucleosynthesis in the various ejecta components. In particular in Fig.~\ref{fig:var_wind} we compare our standard case, in which no europium is produced in the disk wind emerging from a remnant characterized by a long-lived massive NS, to a situation in which all disc winds produce europium, i.e. we assume the same nucleosynthesis with $ 80 \leq A \leq 232$ for all winds. Again for the smallest galaxy case allowing this extra production of europium can increase up to one order of magnitude the quantity of Eu, but we stress that the result is heavily affected by small number fluctuations, with essentially only two realisations (out of 200) showing a significant increase in the retained europium mass. For more massive galaxies the impact is instead milder, determining at most an increase of a factor of a few.

Finally, we have tested the impact of other input parameters of the models,
including the extension of the galaxy in computing the retained mass and the probability of receiving of low kick, i.e. $\xi$ and $p_{l}$. In both cases, the amount of retained material scales as expected in the case of low mass galaxies: an increase from $\xi=3$ to $\xi=5$ translates in an increase of $\sim (5/3)^2$ in the retained mass, due to a more extended disc surface.\footnote{Nevertheless, we expect our standard choice ($\xi=3$, corresponding to three times the radial scale of the exponential profile) to cover a significant fraction of the galaxy in which stars are produced.}
For the latter parameter, a decrease in the probability from 0.7 to 0.6 produces essentially no noticeable differences in the amount of retained mass, except for the least massive galaxy case, where a slightly decrease of the captured mass arises. Any dependence on both $\xi$ and $p_{l}$ becomes less and less relevant for large galaxy masses and, in particular, for a MW-like galaxy.

\begin{figure*}
    \begin{minipage}{0.48\textwidth}
    \includegraphics[width=\textwidth]{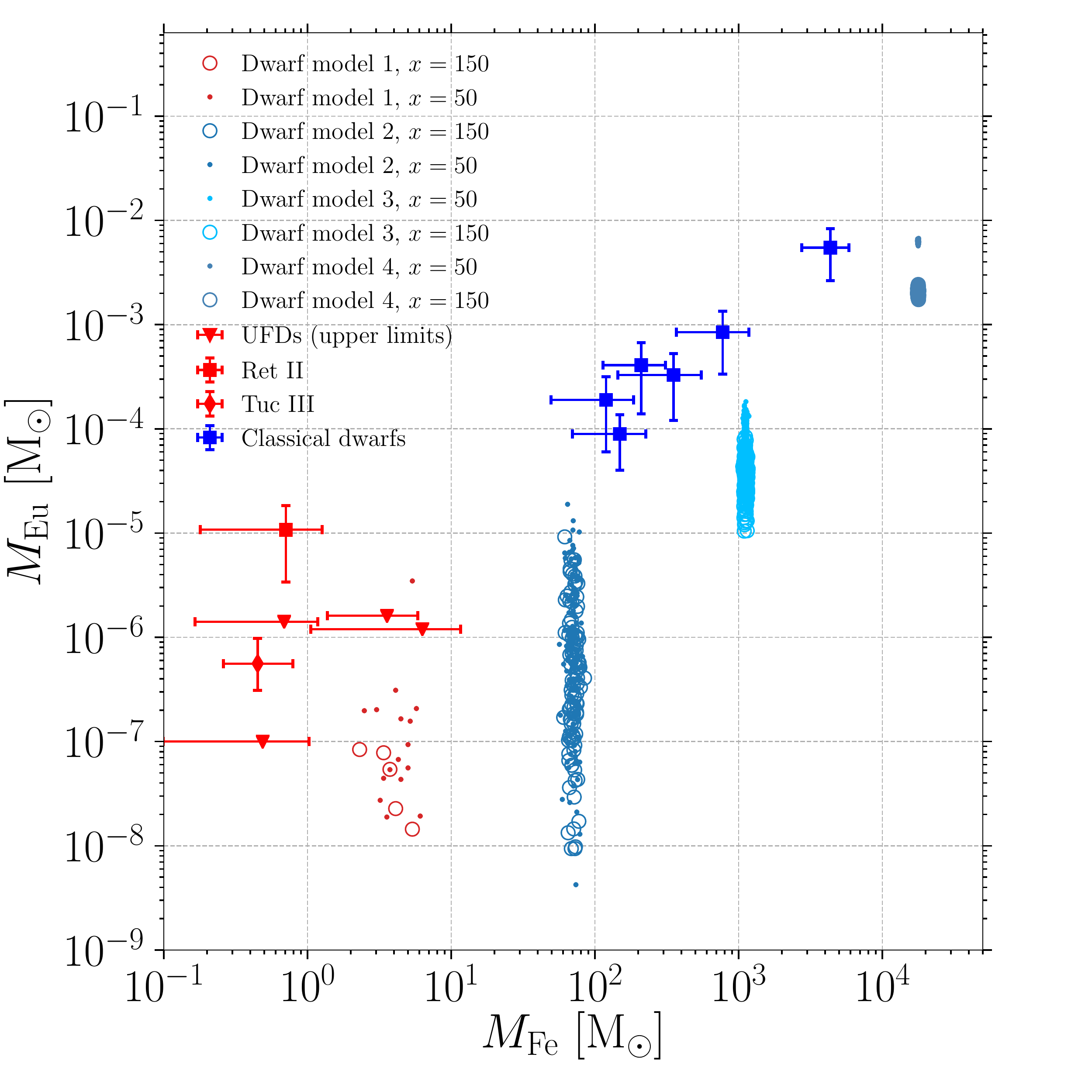} 
    \end{minipage}
    \begin{minipage}{0.48\textwidth}
    \includegraphics[width=\textwidth]{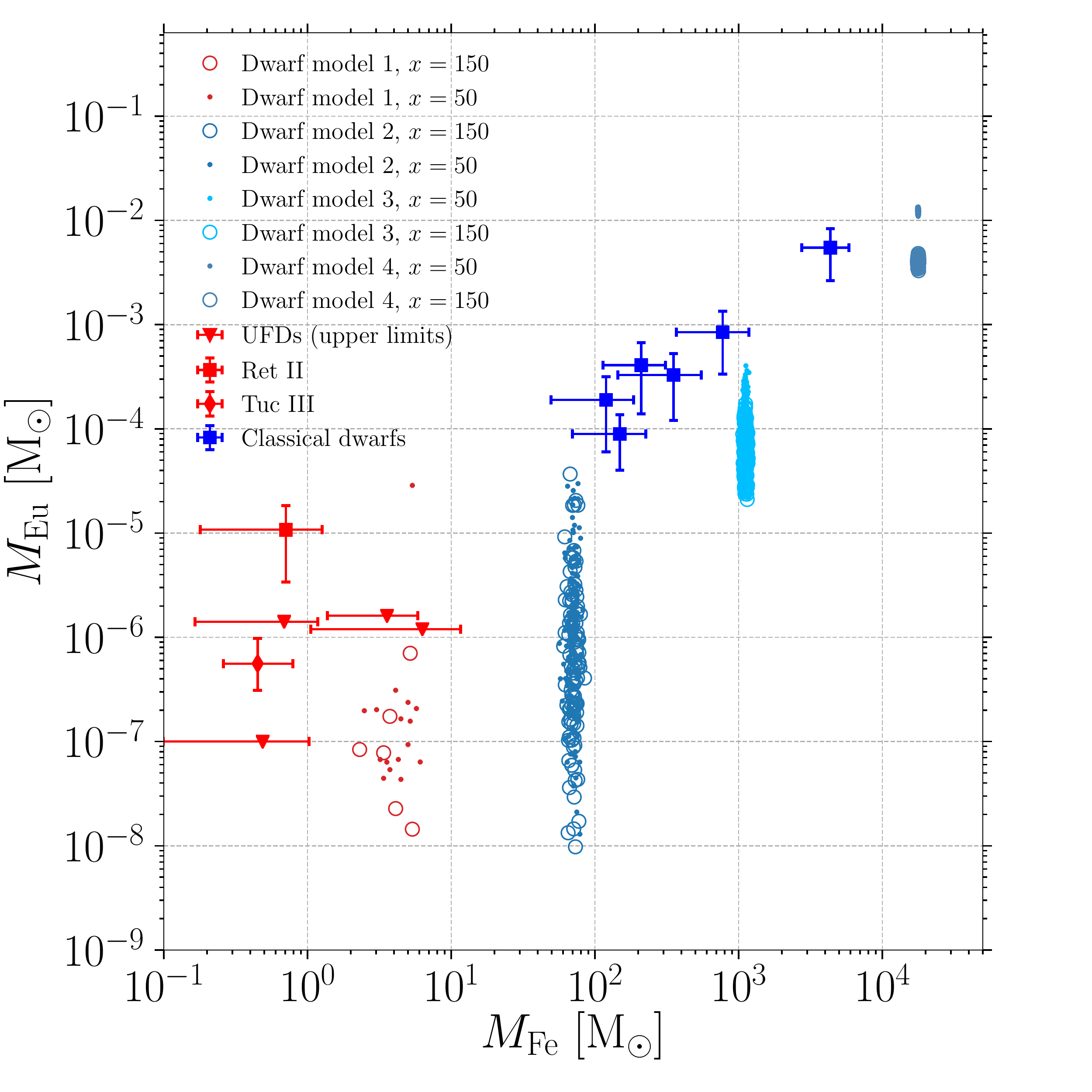}
    \end{minipage}
    \caption{Retained europium mass as a function of iron mass for all our considered dwarf galaxy models. Dots (red for Dwarf model 1, blue for Dwarf model 2 to 4) refer to the optimistic case with $x=50$, while open circles to the fiducial case with $x=150$. On the same figure we also report observational estimates of europium and iron masses for a sample of UFDs (red triangles as upper limits, red square for Reticulum II and red diamond for Tucana III) and classical dwarf (blue squares); we adopted these data from \citet{Beniamini2016b}, apart for Tucana III, for which we calculated europium and iron mass using the same approximations.
    {\it Left panel}: europium production is allowed in dynamical and wind 1 ejecta. {\it Right panel}: europium can be produced in all components (i.e. also wind 2). Despite the large uncertainties, our approach broadly suggests a systematic deficiency in europium at a fixed iron mass.}
    \label{fig:mFe_mEu}
\end{figure*}

\section{Comparison with dwarf galaxy observations}
\label{sec:discussion}

\begin{figure}
    \centering
    \includegraphics[scale=0.36]{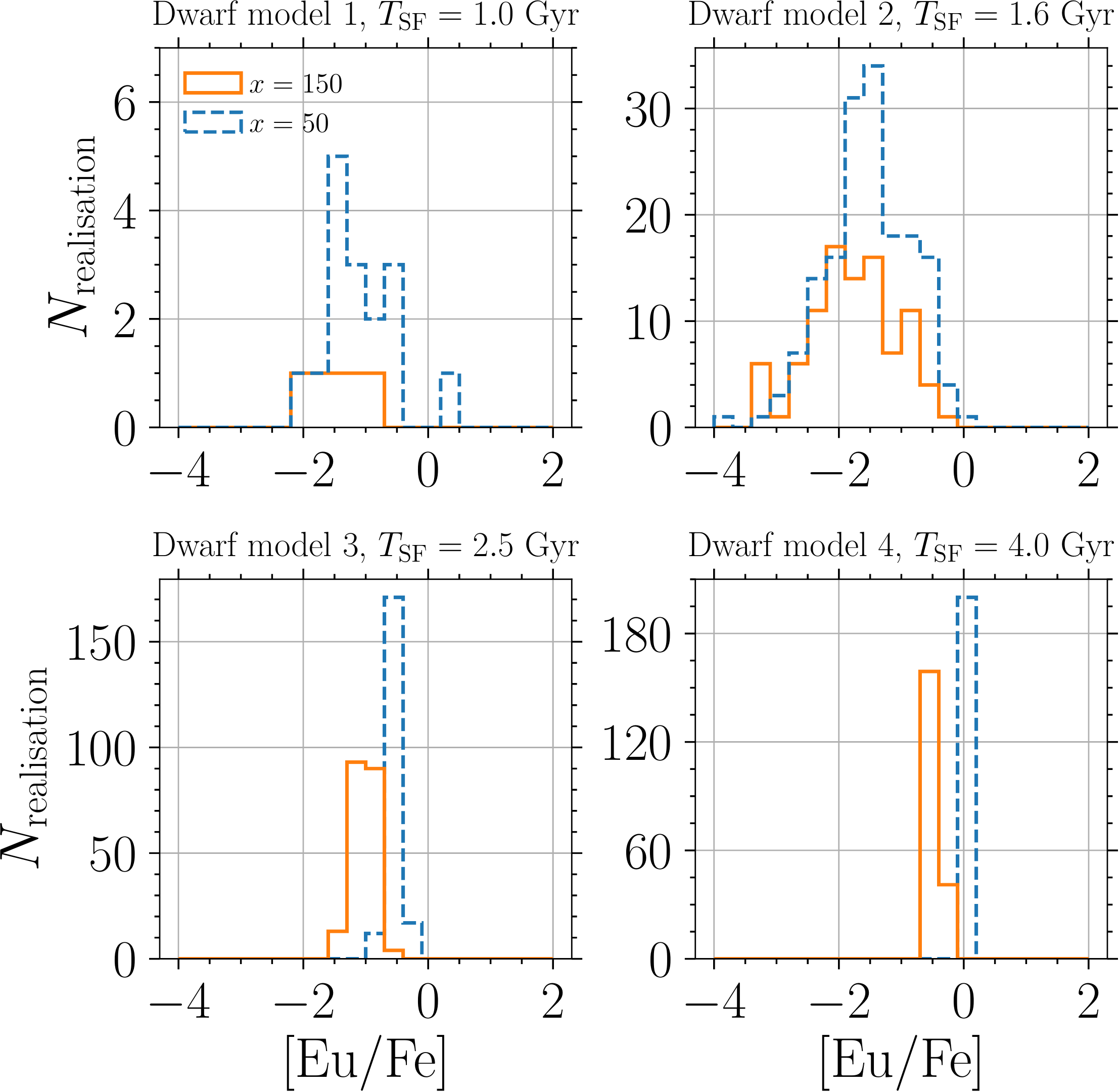}
    \caption{Element abundance of europium over iron as obtained in our models for dwarf galaxies of increasing baryonic mass. Solid orange (dashed cyan) lines refer to a CCSN to NSB rate $x=150$
    ($x=50$).}
    \label{fig:el_abu}
\end{figure}

Fig.~\ref{fig:mFe_mEu} shows the amount of retained europium mass as a function of the iron mass produced by CCSNe explosions. We stress that the nucleosynthesis contribution of SNIa is neglected in our work. This assumption is valid as long as star formation happens significantly before SNIa start to explode (as in the case of UFD galaxies). Otherwise, the amount of iron computed provides a lower limit. In the figure we report the comparison between our results (small dots and empty circles) and observational data points referring to UFDs (red triangles, square for Ret II and diamond for Tuc III) and classical dwarfs (blue square), respectively. Each small dot (empty circle) represents a successful realisation in which europium is effectively captured by the host galaxy assuming a CCSN/NSB ratio of $x = 150$ ($x=50$). The latter values have been chosen by comparing the amount of europium retained in MW-like galaxy models with estimate of europium mass in our Galaxy. Assuming a mass of $5000\msun$ for $r$-process material with $A \geq 90$ \citep[see e.g.][and references therein]{Hotokezaka2018} and an europium mass fraction of $m_{\rm Eu}/m_{r-\rm proc,A \geq 90} = 0.0042$ \citep{Lodders2003}, we found $50 \lesssim x \lesssim 150$, depending on the detailed parameter choice.
Observational data points are taken from figure 1 and table 1 of \citet{Beniamini2016b}. We refer to their section 2 for the presentation of the UFD and classical dwarf samples and for the relevant references.
Concerning Tucana III, we followed the same procedure as in \citet{Beniamini2016b}, but from stellar abundances reported in \citet{Hansen2017}.

From left to right, each group of realisations clusters around specific values of the iron mass, which is proportional to the explored baryonic masses (in the interval $10^5-10^8\msun$), and shows a small dispersion. The europium mass instead shows a much larger scatter, especially for Dwarf model 1 and 2, where variations span three orders of magnitude. On the contrary, at increasing galaxy mass the scatter decrease. Again this result is a combination of the small number of events and the geometric dilution factor.
From Fig.~\ref{fig:mFe_mEu} an increasing trend with galaxy mass for the europium mass is clearly visible, but when compared to the observational data points, our procedure systematically underestimate $m_{\rm Eu}$ for all galaxy models with baryonic mass in the range $10^5-10^7\msun$. To test potential systematic uncertainties in our model, in the right panel of Fig.~\ref{fig:mFe_mEu}, we report the same quantities of the left panel but we consider that europium production is active also in the wind emerging from long-lived remnant. This cause a shift of at most 0.5 dex for $m_{\rm Eu}$, easing the tension, but without definitively solving it. Therefore, despite the large uncertainties that affect the followed procedure, our results even in the most optimistic case (right panel of Fig.~\ref{fig:mFe_mEu}, $x=50$ case), might imply a possible tension with the scenario in which europium enrichment is only generated by NSB mergers.

In Fig.~\ref{fig:el_abu} we provide histograms of $[{\rm Eu/Fe}]$ for all our four dwarf models. Once again, results for models 1,2, and 3 are systematically smaller than the values required to explain abundances in Reticulum II and in classical dwarfs, as reported in table 1 of \citet{Beniamini2016b} by at least one dex, even in the most optimistic case $x=50$.

It is still possible to argue that the effective amount of $r$-process material produced by NSB mergers is not so well constrained and a larger production could solve the problem. However, there is at least another prediction of the model which we found more difficult to reconcile with observations. As previously pointed out, in our model dwarf galaxies show a significant spread of Eu enrichment compared to a relative fixed enrichment of iron. This spread is dependent on the galaxy mass and increasing for less massive objects. Intuitively, this is connected to the shallower potential well that the NSB encounter after the second SN kick in a less massive host galaxy. In the model, we do not follow any possible stochastic enrichment inside the galaxy due to finite dimension of the pollution by our NSMs; the spread obtained is between the average Eu in single galaxies. Therefore, the observational expectation is that classical dwarf galaxies can differ in their average Eu/Fe ratio from 1 dex for the more massive one to more than 3 dex for the lightest ones.
However, according to the data up to now collected, this is not the case. Observed dwarf spheroidal galaxies with similar final stellar masses show a relative small scatter in iron \citep{McConnachie2012}, which is compatible with our model. On the other hand, the enrichment in Eu for classical dwarf is, 
in most of the cases, proportional to that of iron, differently from the
outcome of our numerical modelling. For the classical dwarf, the only hint of a substantial variation of the Eu enrichment is the [Eu/Fe] ratio measured in stars of Sagittarius dwarf galaxy \citep{McWilliam2013}. The situation is different for UFD galaxies.
At the moment most of them seem to have an extremely low enrichment of neutron capture elements with - at present -  the only exception Reticulum II \citep{Ji2016} and possibly Tucana III \citep{Hansen2017}. According to our model, NSB mergers can enrich of the order of 1-2\% galaxies in the mass regime of Reticulum II. Therefore,  we could have simply randomly detected this object, although the chance are relatively low considering the dozen of UFD galaxies with measured stellar abundances.  However, in the case the measurements of europium in four additional stars of Tucana III will be confirmed \citep{Marshall2018}, then the random probability will be certainly too low and a clear tension with the prediction would be confirmed also from this prospective. We should underline that the debate is still on whether UFD galaxies used to be isolated galaxies or are just fractions of larger tidal disrupted objects. This would relax the constrain from this side.

If the stars that we observe in UF and classical dwarfs formed at high redshift $z$, the tension with our models could potentially increase. In fact, we expect the average particle density in the Universe to increase as a function of $z$ as $(1+z)^3$. This implies that for dwarf galaxies forming stars within the first 3.3 Gyr after the Big-Bang (i.e. $z \sim 2$) the IGM could be 10 larger than the present Universe average and likely larger than our standard case ($n_0=10^{-4}{\rm cm^{-3}}$). This increase could limit the extension of the merger remnant, further reducing the dilution factor and, ultimately, the amount of matter enriching the host galaxy.

Our results depends on the chosen binary semi-major axis distribution. Neglecting eccentricity effects, a uniform distribution of $\log_{10}(a)$ translates in a $t^{-1}$ delay time distribution for NSB mergers. \citet{Beniamini.Piran:2019} recently showed that the observed distribution of Galactic NSBs might suggest the presence of a significant population (at least 40\%) of fast merging ($\lesssim 1~{\rm Gyr}$) binaries, larger than the fraction implied by our distribution ($\sim$ 27\%). A closer inspection of the NSB samples used in our models reveals that a significant increase in the enrichment fraction (at least at 10\% level) in most of enriching binaries would require a much larger fraction of binaries merging not only within 1~Gry, but within a few times $10^7$~yr and $10^8$~yr in the $10^5~M_{\odot}$ and $10^7~M_{\odot}$ dwarf galaxy model, respectively.

\section{Conclusions}
\label{sec:conclusions}

NSB mergers are nowadays clearly recognized as one of the major source of $r$-process nucleosynthesis in the Universe and a key player in galactic chemical evolution.
The event rate and the mass per event needed to explain UFD enrichment appears to be consistent with those in the Milky Way and with the NSB merger properties obtained by the analysis of GW170817 \citep[see e.g.][]{Hotokezaka2018}.
In this paper, we have explored the impact of the orbital motion of binary systems of NSs around galaxies prior to merger on the $r$-process enrichment. 
We have found that for low mass systems, i.e. disc galaxies with a baryonic mass $M_{\rm b}$ ranging from $10^5$ up to $10^8\msun$, the motion of the binary due to the kicks imparted by the two SN explosions determines a merger location potentially detached from the disc plane, even for gravitationally bound systems. 

The immediate consequence is a dilution of the amount of $r$-process material retained by the galaxy within its star forming age $T_{\rm SF}$ (and thus potentially available for the next generation of stars).
This effect is more severe for low mass disc galaxies. Assuming a log-flat distribution for the semi-major axis, realistic distributions of NSB parameters, a production rate of one double NS system every 150 CCSNe, and a rather dilute IGM (with a density of $n_0=10^{-4}{\rm cm^{-3}}$), in the least massive case we have explored ($M_{\rm b}=10^5\msun$) a galaxy has a $\sim 10\%$ probability of producing a merging NSB within ($T_{\rm SF} \lesssim 1~{\rm Gyr}$), and a significantly lower probability of retaining $r$-process elements from this single event ($\sim 2\%$). Since the merger happens at distances comparable or larger than the galactic disc size and outside from the disc plane, a fraction ranging between 0.1 and 0.001 of the ejected mass is actually retained (corresponding to $10^{-5}-10^{-4}\msun$) in spite of the ejection of a few $10^{-2}-10^{-3}\msun$ of $r$-process material, most of which stops and mixes with the IGM. This dilution effects is also present in more massive galaxies, but it becomes less and less relevant as $M_{\rm b}$ increases, and it has practically no relevance for a MW-like galaxy.

We have also estimated the amount of Eu retained by the galaxy. We considered that mass ejection in NSB mergers happens through different channels and each channel has a potentially different nucleosynthesis, based on the influence of neutrino irradiation. We have found that the amount of retained Eu is usually $\sim 10^{-3}$ the amount of retained $r$-process material. This is due to the presence of a significant fraction of binaries that produce a long-lived massive NS. For these remnants, discs are usually more massive and the persistent neutrino irradiation suppresses the synthesis of elements between the second and the third $r$-process peaks. The relative fraction of Eu is largely unaffected by the galaxy mass, however the paucity of enrichment events in the low mass case produces larger fluctuations.

We have tested the robustness of our results with respect to the parameters entering the model. The most relevant parameters are the fraction of double NS systems with respect to the number of CCSNe ($x$) and the IGM density ($n_0$). For the former, even considering the most optimistic case (i.e. one double NS system every 50 CCSNe), the results stay within a factor of a few our standard case. For the latter, a significantly larger density (which is what we expect in the early Universe) determines less extended NSB remnants and prevents $r$-process enrichment in the least massive case.

The precise modelling of the composition of NSB ejecta is still affected by uncertainties, as well as our knowledge of the fraction of systems that forms a long-lived remnant. Thus, we have repeated our analysis assuming that all disc winds produce Eu in the same ratio. We verified that our standard results are qualitatively robust, since the amount of retained Eu increases only by a factor of few. 
Calculations of the retained $r$-process material allows us to compare with observed abundances of iron and europium from UFD and classical dwarfs. Due to the dilution effect on the retained $r$-process material, all our realizations show a systematic deficiency in the europium abundance compared with observed abundances. Additionally, for a fixed galaxy model and especially for low mass galaxies, the large intrinsic variability introduced by the dilution process and small number statistics is not visible in the observations. Since the dilution is practically negligible in the MW-like galaxy model, in that case we have estimated the CCSN to NSB rate within our assumptions and we found $50 \lesssim x \lesssim 150$, with a preference on the low value side.

Both the discrepancies on the absolute values and on their spread could potentially imply a tension in the explanation of the europium production in dwarf galaxies as produced by NSB mergers only. This conclusion could depend on observational uncertainties in the estimate of the elemental abundances or it could imply a significant change in one or more of the canonical assumed parameters. 
However, to reconcile our models with observations a systematically optimistic choice of parameters is required, together with a significant revision of the observed elemental abundances.

On the one hand, a possible solution to the discrepancy is to assume the existence of a fast merger population \citep[e.g.][]{Beniamini.Piran:2019}, such that a significant fraction of NSBs merge within $10^8$~yr, or even a few times $10^7$~yr. Under these conditions the merger does not happen too far from the galaxy and the dilution factor becomes $\gtrsim 0.1$.
If, on the other hand, this tension is confirmed, the existence of additional sites for the production of $r$-process elements is necessary. A similar conclusion was also obtained by other authors and motivated by the difficulty to inject r-process elements early enough to explain the Eu abundances in metal-poor stars \citep{Matteucci2014,Cescutti2015,Wehmeyer2015,Haynes2019}. Moreover, also \citet{Cote2019} and \citet{Simonetti2019} further suggested an extra production site of europium as a possible way of reproducing the decreasing trend of [Eu/Fe] in the Galactic disk, a different feature that is the result of $\sim12$ Gyr of chemical evolution.

BH-NS mergers are also possible sources of $r$-process elements \citep[e.g.][]{Shibata2011}.
The lack of a MNS in the remnant favours the production of the heaviest $r$-process elements \citep[e.g.][]{Roberts2017,Lippuner2017}. This could enhance the amount of retained europium by a factor of a few. However, even if BH-NS merger rates are highly unconstrained, we expect them to be significantly lower than NSB merger rates \citep{Abbott2018LRR}. A possible solution would thus require qualitatively different binary parameters, for example much lower kick velocities or much smaller initial separations, resulting from a possibly different binary evolution. This could however be in tension with population synthesis results, considering that a significant amount of ejecta from BH-NS merger requires a not too large mass ratio ($\lesssim 5$), for moderately high BH spins.

An alternative solution is represented by special classes of CCSNe, able to produce $r$-process elements with an amount comparable to the one of NSB mergers and with a similar rate. Possible examples include magnetically-driven CCSNe \citep{Fujimoto2008,Winteler2012,Nishimura2015,Moesta2018} and disk ejecta from collapsar models \citep[e.g.][]{Malkus2012,Siegel2019}. In these cases the ejection of $r$-process material happens still on a sufficiently short time scale and inside the galaxy, such that no significant dilution factor affects the enrichment \citep{Beniamini2018}. Magnetically-driven CCSNe have also the advantage that they could also explain the chemical enrichment of Galactic halo \citep{Cescutti2014}.
Further multi-physics studies combining galactic (chemical) evolution, binary population synthesis, NSB mergers, and taking into account nuclear uncertainties are required to address all the open issues in the field. In this respect, a crucial role is represented by the forthcoming determination of more precise compact binary merger rates by the Advanced LIGO and Virgo detectors \citep{Aasi2015,Acernese2015,Abbott2018LRR}. 

\section*{Acknowledgements}

MB, AP and MD acknowledge CINECA, under the TEONGRAV initiative, for the availability of high performance computing resources and support. G. Cescutti acknowledges financial support from
the European Union Horizon 2020 research and innovation programme
under the Marie Sk\l odowska-Curie grant agreement No. 664931.



\bibliographystyle{mnras}
\bibliography{biblio} 






\bsp	
\label{lastpage}
\end{document}